\documentclass[%
preprint,
nofootinbib,
 amsmath,amssymb,
twocolumn,
10pt
]{revtex4-1}
\usepackage{appendix}
\usepackage[bottom]{footmisc}
\usepackage[utf8]{inputenc}
\usepackage{slashed}
\usepackage[
colorlinks=true,
linkcolor=blue,
breaklinks=true,
urlcolor=blue,
citecolor=blue]{hyperref}
\usepackage{multirow}
\usepackage{diagbox}
\usepackage{graphicx}
\usepackage{dcolumn}
\usepackage{bm}
\usepackage{hyperref}
\usepackage{setspace}
\usepackage{makecell}
\usepackage{threeparttable}
\usepackage{footnote}
\usepackage{tikz}
\usepackage{tikz-feynman}
\usepackage{float}
\allowdisplaybreaks[4]

\newcommand{\bra}[1]{\langle #1 |}
\newcommand{\ket}[1]{| #1 \rangle}
\newcommand{\bracket}[2]{\langle #1 |#2\rangle}

\newcommand{\lra}[1]{\left(#1\right)}

\tikzfeynmanset{
every vertex/.style={red, dot},
every particle/.style={blue},
every blob/.style={draw=black!40!black, pattern color=black!40!black},
}
\usetikzlibrary{decorations.markings}
\usetikzlibrary{snakes}
\tikzset{
 photon/.style={decorate, decoration={snake}, draw=black},
    electron/.style={draw=black, postaction={decorate},
        decoration={markings,mark=at position .55 with {\arrow[draw=black]{>}}}},
    gluon/.style={decorate, draw=magenta,
        decoration={coil,amplitude=3pt, segment length=4pt}},
    scalar/.style={dashed,line width=.6pt, postaction={decorate},
        decoration={markings,mark=at position .55 with {\arrow[draw=black]{>}}}},
}

\begin{document}

\title{Prediction of possible $DK_1$ bound states }

\author{Xiang-Kun Dong$^{1,2}$}
 \ \email{dongxiangkun@itp.ac.cn}
\author{Bing-Song Zou$^{1,2,3}$}
\ \email{zoubs@itp.ac.cn}

\address{%
$^1$ CAS Key Laboratory of Theoretical Physics, Institute of Theoretical Physics,\\
Chinese Academy of Sciences, Beijing 100190, China\\
$^2$ School of Physical Sciences, University of Chinese Academy of Sciences, Beijing 100049, China\\
$^3$ School of Physics, Central South University, Changsha 410083, China
}


\begin{abstract}
Stimulated by recent experimental observation of $X_1(2900)$ just below the $\bar D_1K$ threshold, we extend our previous study of $\bar D_1D$ S-wave bound state by vector meson exchange to $\bar D_1K$ system as well as similar $\bar DK_1$, $D_1K$ and $DK_1$ systems to look for possible bound states. We find that the potential of $D K_1$ is attractive and strong enough to form bound states with mass around 3110 MeV for $DK_1(1270)$ and 3240 MeV for $DK_1(1400)$. $D_1 K$ is also attractive but weaker, hardly enough to form bound states. While $\bar DK_1$ becomes further less attractive, the potential between $\bar D_1 K$ is the weakest, definitely too weak to form any bound state, which excludes the recently observed $X_1(2900)$ to be a $\bar D_1 K$ bound state. We also give the decay properties of the predicted $D K_1$ bound states.
\end{abstract}

\maketitle
\section{Introduction}

Two new open flavor states were recently reported by LHCb collaboration~\cite{Aaij:2020ypa} in the $D^-K^+$ final state in $B^{+} \rightarrow D^{+} D^{-} K^{+}$ with statistical significance much larger than 5$\sigma$. The fitted masses and widths are
\begin{align*}
X_{0}(2900): &M=2866.3 \pm 6.5 \pm 2.0 \mathrm{MeV} / \mathrm{c}^{2} \\
&\Gamma=57.2 \pm 12.2 \pm 4.1 \mathrm{MeV} / \mathrm{c}^{2} \\
X_{1}(2900): &M=2904.1 \pm 4.8 \pm 1.3 \mathrm{MeV} / \mathrm{c}^{2} \\
&\Gamma=110.3 \pm 10.7 \pm 4.3 \mathrm{MeV} / \mathrm{c}^{2}
\end{align*}
with corresponding quantum numbers of $J^{P}=0^+$ and $1^-$, respectively. They are very interesting and of great importance since if confirmed to be real resonances instead of kinetic effects, each of them consists of at least four (anti)quarks which are beyond the conventional quark model. Up to now dozens of works have made efforts to understand these two states~\cite{Karliner:2020vsi,He:2020jna,Zhang:2020oze,Wang:2020xyc,Lu:2020qmp,Chen:2020aos,Hu:2020mxp,Liu:2020nil,He:2020btl,Huang:2020ptc,Xue:2020vtq,Molina:2020hde,Agaev:2020nrc,He:2020btl,Liu:2020orv,Burns:2020epm,Albuquerque:2020ugi,Chen:2020eyu,Mutuk:2020igv,Burns:2020xne}. On one hand, $X_{0}(2900)$ is explained as a $cs\bar u\bar d$ compact tetraquark~\cite{Karliner:2020vsi,He:2020jna,Zhang:2020oze,Wang:2020xyc,Wang:2020prk}, but this explanation is disfavored by explicit calculation of spectra using extended quark model~\cite{Lu:2020qmp}. On the other hand, $X_{0}(2900)$ is also regarded as a molecule of $\bar D^*K^*$~\cite{Chen:2020aos,Hu:2020mxp,Liu:2020nil,Huang:2020ptc,He:2020btl,Molina:2020hde,Agaev:2020nrc,Mutuk:2020igv,Xiao:2020ltm}. It worth mentioning that in Ref.~\cite{Molina:2010tx} a bound state of $D^*\bar K^*$ with $I=0,\ J^P=0^+$ whose mass is close to that of $X_0(2900)$ is predicted.  Similarly, $X_{1}(2900)$ is explained as a compact tetraquark~\cite{Chen:2020aos,He:2020jna,Xue:2020vtq,Molina:2020hde,Agaev:2020nrc,Mutuk:2020igv,Agaev:2021knl}, a $\bar D_1K$ virtual state~\cite{He:2020btl} or a $\bar D_1K$ bound state~\cite{Tan:2020cpu,Qi:2021iyv}. Besides, these exotic signals in LHCb's observation are explained as kinetic effect, namely, triangle singularity in Refs.\cite{Liu:2020orv,Burns:2020epm}. The conflicts among these different interpretations suggest that more experimental results are needed to pin down the nature of these states.

Since the discovery of $\chi_{c0}(3872)$~\cite{Choi:2003ue} in 2003, plenty of exotic states or candidates are observed experimentally, many of which are close to certain hadron pair thresholds and are explained as hadronic molecules~\cite{Guo:2017jvc}. Note that $X_{0}(2900)$ and $X_1(2900)$ are about 30 and 10 MeV below $\bar D^*K^*$ and $\bar D_1 K$ threshold, respectively. Therefore, it is natural to explore the existence of $\bar D^*K^*$ and $\bar D_1 K$ molecules, which have already been explored in Refs.~\cite{Chen:2020aos,Hu:2020mxp,Liu:2020nil,He:2020btl,Mutuk:2020igv} with different methods. In our previous works~\cite{Dong:2019ofp,Dong:2021juy} we applied the meson exchange model to $DD_1$ system to investigate the molecular explanation of $Y(4260)$ and an exotic $1^{-+}$ molecule has been predicted.  We now extend the study to $DK_1$ and $D_1K$ systems to look for possible bound states. 

Actually, the states related to $c\bar s$ have attract attention since the discovery of $D_{s0}^*(2317)$~\cite{Aubert:2003fg,Besson:2003cp} which is explained as a $DK$ molecule. See Ref.~\cite{Chen:2016spr} for a review on the spectrum of $D_{sJ}$ mesons. Besides the plenty of phenomenological researches, the direct and indirect calculations on lattice~\cite{Liu:2012zya,Torres:2014vna,Bali:2017pdv,Cheung:2020mql} provide strong evidences favoring the molecular explanation of $D_{s0}^*(2317)$, which almost settles the dispute on the nature of $D_{s0}^*(2317)$.  Based on this assumption, more kaonic and charmed meson molecules, including $D_1K$, are predicted, which are possible interpretations of some observed $D_{sJ}$ states~\cite{Guo:2011dd}.

This work is organized as follows. In section II, the potentials of different systems are derived and the binding energy of possible bound states are calculated. In section III, the decay patterns of the predicted bound states are estimated. A brief summary is given in section IV.

\section{The potentials and possible bound states}
\begin{figure}[h]
\centering
\includegraphics{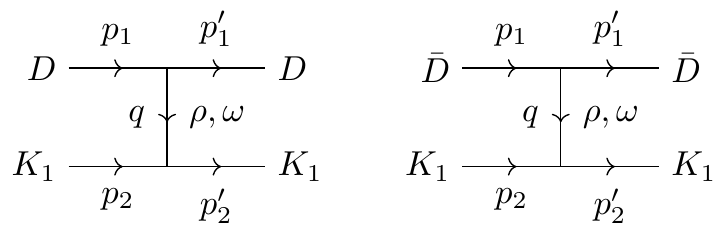}

\includegraphics{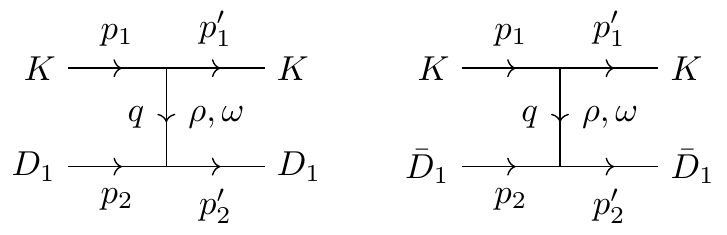}

\caption{Feynman diagrams for vector meson exchange between $D K_1$, $\bar D K_1$, $D_1 K$ and $\bar D_1 K$.}\label{fig:DK1feyn}
\end{figure}

The meson exchanged interactions for $DK_1$ and $D_1K$ systems are illustrated in Fig.(\ref{fig:DK1feyn}) respectively. The relevant Lagrangians are collected in the following.

The couplings of heavy mesons and light vector meson can be described by the effective Lagrangians with the hidden gauge symmetry~\cite{Casalbuoni:1996pg}. For $D$ and $D_1$ mesons, the Lagrangians read explicitly~\cite{Ding:2008gr} 

\begin{align}
\mathcal{L}_{\mathrm{DDV}}&=\frac{i\beta g_{V}}{\sqrt{2}} \left(\mathrm{D}_{b} \stackrel{\leftrightarrow}{\partial}_{\mu} \mathrm{D}_{a}^{\dagger} - \overline{\mathrm{D}}_{a} \stackrel{\leftrightarrow}{\partial}_{\mu} \overline{\mathrm{D}}_{b}^{\dagger}\right)V_{b a}^{\mu}\label{eq:LDDV}\\
\mathcal{L}_{\mathrm{D}_{1} \mathrm{D}_{1} V}&=\frac{i\beta_2 g_{V}}{\sqrt{2}} \left(\mathrm{D}_{1 b}^{\nu} \stackrel{\leftrightarrow}{\partial}_{\mu} \mathrm{D}_{1 a \nu}^{\dagger}-\overline{\mathrm{D}}_{1 a \nu} \stackrel{\leftrightarrow}{\partial}_{\mu} \overline{\mathrm{D}}_{1 b}^{\nu \dagger}\right) V_{b a}^{\mu}\label{eq:LD1D1V}
\end{align}
where
\begin{align}
D_{(1)}&=(D_{(1)}^0,D_{(1)}^+,D_{s(1)}^+)\\
V&=\left(\begin{array}{ccc}{\frac{1}{\sqrt{2}} \rho^{0}+\frac{1}{\sqrt{2}}} \omega & {\rho^{+}} & {K^{*+}} \\ {\rho^{-}} & {-\frac{1}{\sqrt{2}} \rho^{0}+\frac{1}{\sqrt{2}} \omega} & {K^{* 0}} \\ {K^{*-}} & {\bar{K}^{* 0}} & {\phi}\end{array}\right)
\end{align}
with $g_V=5.8$~\cite{Bando:1987br} and $\beta=0.9$~\cite{Isola:2003fh}. As analysed in Ref.~\cite{Dong:2019ofp}, $\beta_2\approx -\beta$. Here we have taken the ideal mixing between $\omega_0$ and $\omega_8$. 

In the light meson sector, the nature of $K_1(1270)$ and $K_1(1400)$ that both are conventional $q\bar q$ mesons~\cite{Burakovsky:1997dd, Suzuki:1993yc, Cheng:2003bn, Yang:2010ah, Hatanaka:2008xj, Tayduganov:2011ui, Divotgey:2013jba, Zhang:2017cbi} or $K_1(1270)$ is a dynamically generated resonance with possible two pole structure ~\cite{Roca:2005nm, Geng:2006yb, Wang:2019mph}, is still controversial and in this work we adopt the former one. Due to the mass difference between $u/d$ and $s$ quarks, the axialvector $K_{1A}$ ($^3P_1$ state) and pseudovector $K_{1B}$ ($^1P_1$ state) mix and generate the two physical resonances $K_1(1270)$ and $K_1(1400)$. Following Ref.~\cite{Divotgey:2013jba} the mixing is parameterized as

\begin{align}
    \left(\begin{array}{cc}{\left|K_{1}(1270)\right\rangle}  \\ {\left|K_{1}(1400)\right\rangle}\end{array}\right)=\left(\begin{array}{cc}{\cos \phi} & {-i\sin \phi}\\ {-i\sin \phi} & {\cos \phi} \end{array}\right)\left(\begin{array}{c}{\left|K_{1A}\right\rangle} \\ {\left|K_{1B}\right\rangle}\end{array}\right)\label{eq:mixing}
\end{align}
with the mixing angle $\phi$ determined to be $(56.4\pm4.3)^{\circ}$.  

From chiral perturbation theory, the pseudoscalar-pseudoscalar-vector coupling reads
\begin{align}
\mathcal{L}_{\rm{PPV}}&=i\sqrt{2}\,G_{\rm V}\,{\rm{Tr}}\lra{[\partial_{\mu}P,P]V^{\mu}}.\label{eq:Lagran_PPV}
\end{align}
and analogously,
\begin{align}
\mathcal{L}_{\rm{AAV}}&=i\sqrt{2}\,G_{\rm V}'\,{\rm{Tr}}\lra{[\partial_{\mu}A^\nu,A_\nu]V^{\mu}}\label{eq:Lagran_AAV}\\
\mathcal{L}_{\rm{BBV}}&=i\sqrt{2}\,G_{\rm V}'\,{\rm{Tr}}\lra{[\partial_{\mu}B^\nu,B_\nu]V^{\mu}}\label{eq:Lagran_BBV}
\end{align}
with
\begin{align}
&P=\left(\begin{array}{ccc}{\frac{1}{\sqrt{2}} \pi^{0}+\frac{1}{\sqrt{6}} \eta_{8}} & {\pi^{+}} & {K^{+}} \\ {\pi^{-}} & {-\frac{1}{\sqrt{2}} \pi^{0}+\frac{1}{\sqrt{6}} \eta_{8}} & {K^{0}} \\ {K^{-}} & {\bar{K}^{0}} & {-\frac{2}{\sqrt{6}} \eta_{8}}\end{array}\right),\\ 
&A(B)=\left(\begin{array}{ccc}{*} & {*} & {K_{1A(B)}^{+}} \\ {*} & {*} & {K_{1A(B)}^{0}} \\ {K_{1A(B)}^{-}} & {\bar{K}_{1A(B)}^{0}} & *\end{array}\right).
\end{align}
The coupling constant $G_{\rm V}\approx3.0$ was estimated from the decay width of $\rho\to\pi\pi$~\cite{Zhang:2006ix} and we adopt the approximation $G'_{\rm V}\approx -G_{\rm V}$. Note that the irrelevant axialvector and pseudovector states are not shown explicitly in the corresponding multiplet and represented by ``*". Expanding Eq.(\ref{eq:Lagran_PPV}) we obtain the couplings between $KKV$,
\begin{align}
\mathcal{L}_{\rho K K}&=i G_{V}\left[\bar{K} \vec{\tau}\left(\partial_{\mu} K\right)-\left(\partial_{\mu} \bar{K}\right) \vec{\tau} K\right] \cdot \vec{\rho}^{\mu},\label{eq:KKrho}\\
\mathcal{L}_{\omega K K}&=i G_{V}\left[\bar{K}\left(\partial_{\mu} K\right)-\left(\partial_{\mu} \bar{K}\right) K\right] \omega^{\mu},\label{eq:KKomega}\\
\mathcal{L}_{\phi K K}&=-\sqrt{2} i G_{V}\left[\bar{K}\left(\partial_{\mu} K\right)-\left(\partial_{\mu} \bar{K}\right) K\right] \phi^{\mu}\label{eq:KKphi}
\end{align}
with
\begin{align}
K&=\left(\begin{array}{c}{K^{+}} \\ {K^{0}}\end{array}\right), \bar{K}=\left(K^{-}, \bar{K}^{0}\right),\\ \vec{\rho}&=\lra{\frac{\rho^++\rho^-}{\sqrt{2}},\frac{\rho^--\rho^+}{i\sqrt{2}},\rho^0}
\end{align}
and $\vec{\tau}$ the Pauli matrices in isospin space. The coupling between $K_1K_1V$ has the same form as Eqs.(\ref{eq:KKrho},\ref{eq:KKomega},\ref{eq:KKphi}).

The potentials in momentum space are 
\begin{equation}
\tilde V({\bf q})=f_{\mathrm I}g_Dg_K\frac{1}{|{\bf q}|^2+m_{\rm V}^2}\label{eq:Poten_Momentum}
\end{equation}
where the involved constants are listed in Tab.(\ref{tab:conts}).  The relative signs of $g_D$ and $g_K$ for different systems are determined by the following facts. It is easy to see that the vector meson exchange yields an attractive potential of $D\bar D$ system and thus $DK$ system is also attractive since $c$ and $s$ quarks are spectators during the interaction. These are consistent with the potentials from the Weinberg-Tomozawa term~\cite{Weinberg:1966kf, Tomozawa:1966jm}. Therefore, $D\bar D_1$, $DK_1$ and $D_1 K$ should all have attractive interactions~\cite{Guo:2011dd}, which in turn fixes the potentials of all other related systems.  These determined signs are in agreement with the fact that the $Y(4260)$ contains sizeable components of $D\bar D_1$ bound state~\cite{Wang:2013cya,Chen:2019mgp}.

\begin{table}
\caption{Constants in Eq.(\ref{eq:Poten_Momentum}) for different systems. $f_I$ is the isospin factors with $I=0$ or 1. $\rho$ and $\omega$ are the exchanged particles. $g_{D/K}$ stands for the $D_{(1)}D_{(1)}V/K_{(1)}K_{(1)}V$ coupling constants.}\label{tab:conts}
\begin{spacing}{1.4}
\begin{tabular}{|c|c|c|c|c|c|c|}
\hline
\multirow{2}{*}{} & \multicolumn{2}{c|}{$f_0$} & \multicolumn{2}{c|}{$f_1$} & \multirow{2}{*}{$g_D$}           & \multirow{2}{*}{$g_K$} \\ \cline{2-5}
                        & \ \ $\rho$\ \        &\ \  $\omega$    \ \   & \ \ $\rho$  \ \      & \ \ $\omega$  \ \   &                               &                     \\ \hline
$DK_1$               & 3         & 1          & -1        & 1          & $\beta g_V/{\sqrt2}$  & $-G_V$                       \\ \hline
$D\bar{K}_1$            & -3        & 1          & 1         & 1          & $\beta g_V/{\sqrt2}$  &$G_V$                                        \\ \hline
$D_1K$               & 3         & 1          & -1        & 1          &$-\beta g_V/{\sqrt2}$ & $G_V$                \\ \hline
$D_1\bar K$              & -3        & 1          & 1         & 1          & $-\beta g_V/{\sqrt2}$ & $-G_V$                                   \\ \hline
\end{tabular}
\end{spacing}
\end{table}

A form factor should be introduced at each vertex to account for the finite size of the involved mesons. Here we take the commonly used monopole form factor
\begin{equation}
F(q,m,\Lambda)=\frac{\Lambda^2-m^2}{\Lambda^2-q^2},\label{eq:form_factor}
\end{equation}
which in coordinate space can be looked upon as a spherical source of the exchanged mesons~\cite{Tornqvist:1993ng}. The potentials in coordinate space can be obtained by Fourier transformation of Eq.(\ref{eq:Poten_Momentum}), together with the form factor Eq.(\ref{eq:form_factor}),

\begin{align}
V\left(\mathbf{r}, m_{{\rm{V}}}\right)=f_{\mathrm I}\,g_D\,g_K{\Big(}&m_{\rm{V}}Y(m_{\rm{V}} r) -\Lambda Y(\Lambda r)\notag\\
&-\frac{1}{2}(\Lambda^{2}-m_{\rm{V}}^{2})r Y(\Lambda r)\Big)\label{eq:Poten_spatial}
\end{align}
with $Y(x)=e^{-x}/4\pi x$ the Yukawa potential.

\begin{figure}
\centering

\includegraphics[width=\linewidth]{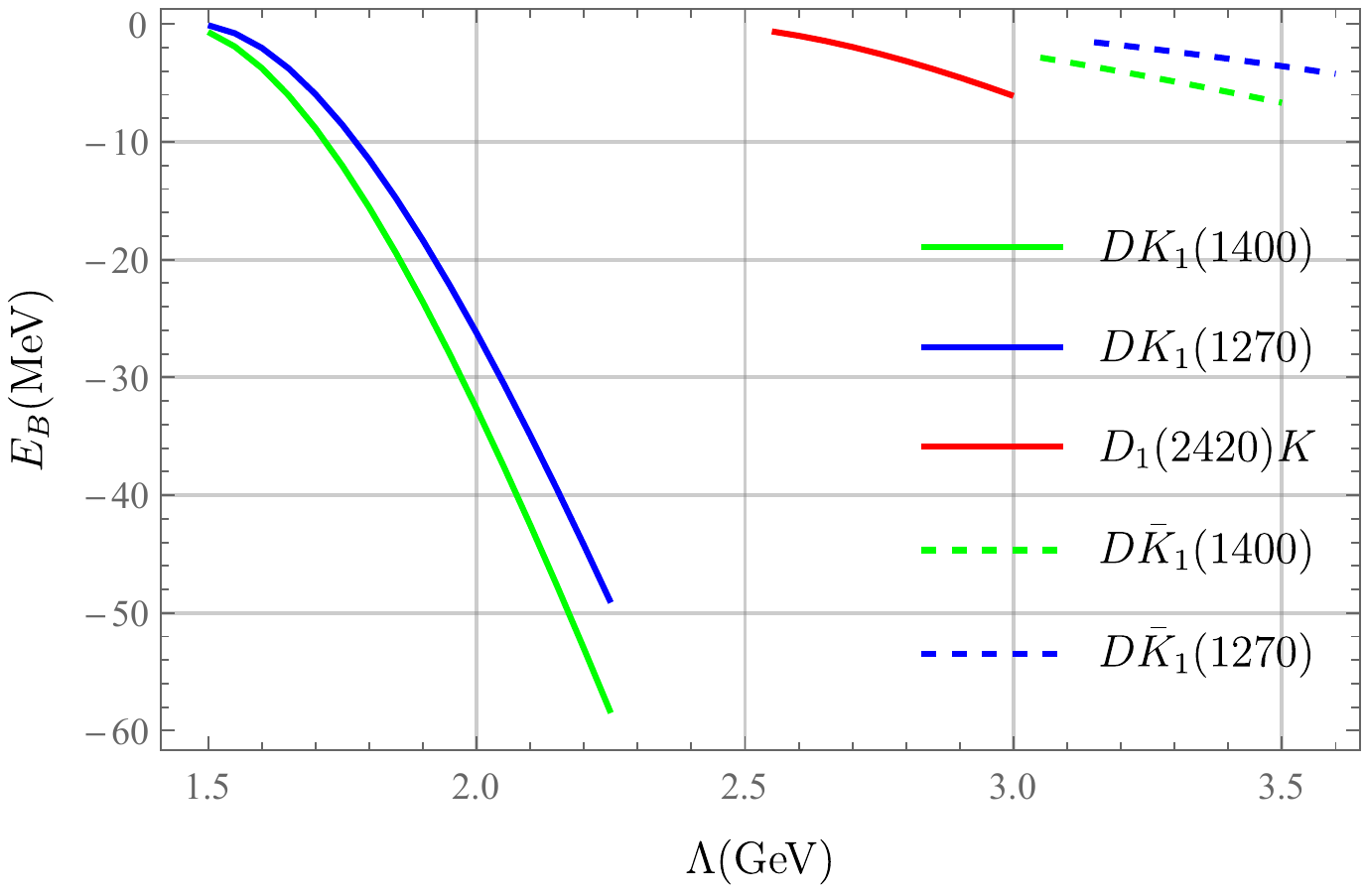}
\caption{Binding energies v.s. cutoff $\Lambda$ for different isoscalar systems.}\label{fig:BindingE}

\end{figure}

The Schr\"odinger equations for the potentials in Eq.(\ref{eq:Poten_spatial}) can be solved numerically. Let's first focus on the isovector system. For $DK_1$ or $D_1K$ system, $\rho$ and $\omega$ exchanges have opposite contributions, leading to an almost vanishing potential due to the their degenerated masses, while for $D\bar K_1$ or $D_1\bar K$ system, both  $\rho$ and $\omega$ exchanges yield repulsive potentials. Therefore, no isovector bound states are possible. Isoscalar systems are easier to form bound states. For $DK_1$ or $D_1K$ system, both  $\rho$ and $\omega$ exchanges yield attractive potentials. For $D\bar K_1$ or $D_1\bar K$, the attractive potential from $\rho$ exchange is about three times of the repulsive potential from $\omega$ exchange, resulting in a total attractive potential that is about 1/2 of that of $D K_1$ or $D_1 K$ systems. The binding energies of different systems from numerical calculation are shown in Fig.(\ref{fig:BindingE}). We can see that when $\Lambda\gtrsim 1.5$ GeV, $DK_1$ can form bound states while for $D_1K$ a much larger $\Lambda\gtrsim 2.6$ GeV is needed due to the smaller reduced mass. For the $D\bar K_1$ system, bound states are possible only when $\Lambda\gtrsim 3.0$ GeV, which is far beyond the empirical region of the cutoff, say $1\sim 2$ GeV. We fail to find any bound states of $D_1\bar K$ with any $\Lambda$ value. It is worth mentioning that the $X_1(2900)$ with $J^{P}=1^{-}$ might be related to $\bar D_1 K$ system since they can couple in S-wave and the mass of  $X_1(2900)$ is just 10 MeV below the threshold of $\bar D_1 K$ but our result excludes the possibility of explaining  $X_1(2900)$ as a $\bar D_1 K$ molecule.

The interactions we considered above also work in the $DK$ system. The parameters that produce a $DK$ bound state corresponding to $D_{s0}(2317)$ can also result in a $D_1 K$ bound state with a similar binding energy, but cannot bind $\bar D_1 K$ together.

These calculation can be easily extended to the $B_1K$ and $BK_1$ cases. The interactions should be the same due to the heavy quark symmetry. For $B_1 K$, the reduced mass is almost the same as that of $B_1 K$ and hence a large cutoff $\Lambda\gtrsim 2.6$ GeV is needed to produce bound states. While for $BK_1$, the reduced mass gets larger than that of $DK_1$ and hence deeper bound states than $DK_1$ are expected. For example, the binding energy of $DK_1(1400)$ and $BK_1(1400)$ are $15$ and $50$ MeV when $\Lambda=1.8$ GeV.

\section{The decay properties of $DK_1$ molecules}
\begin{figure}[h]
\centering
\includegraphics[width=\linewidth]{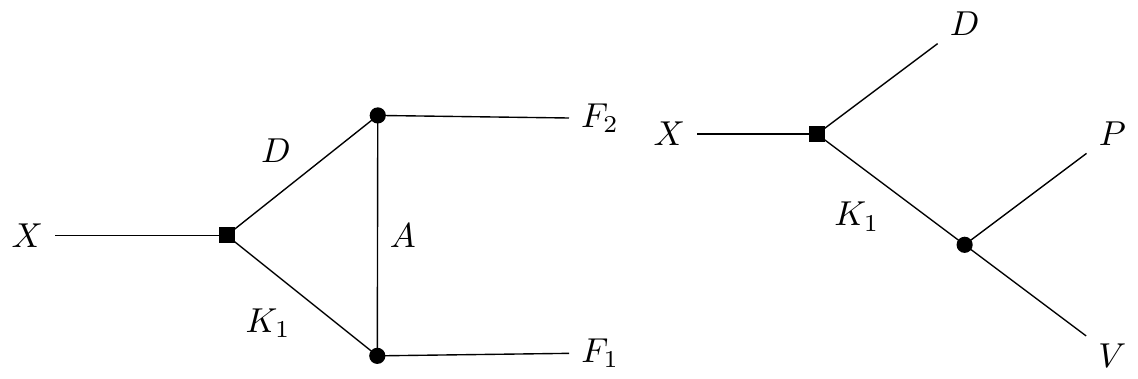}
\caption{Two-body and three-body decays of molecules composed of $DK_1$.}\label{fig:decayfeyn}
\end{figure}
With reasonable cutoff we obtain the possible bound states of $DK_1(1400)$ and $DK_1(1270)$, denoted by $X$, whose binding energies lie in the range of $0\sim$ 30 MeV. It is now desirable to estimate the decay patterns of the predicted molecular states to provide some guidance of their experimental search. We assume that such molecules decay through their components as illustrated in Fig.\ref{fig:decayfeyn}. Since $K_1$'s have quite large decay widths, we assume that the three-body decays of molecules are dominated by the one shown in Fig.\ref{fig:decayfeyn} where $P,V=\pi, K^*$ or $K,\rho/\omega$. All possible two-body strong decay channels are listed in Table~\ref{tab:channels}. Note that some exchanged particles in this table are marked by underlines because these diagrams are expected to have much smaller contributions than others. Therefore, we do not consider these diagrams for simplicity.
\begin{table}
	\centering
	\caption{\label{tab:channels}Two-body decay channels for the $DK_1$ molecule states considered in our calculation. Diagrams with heavier exchanged particles marked by underlines are much smaller and dropped in our calculation.}
	\begin{threeparttable}
\begin{spacing}{1.4}
			\begin{tabular}{c|c}
				\Xhline{1.0pt}
				 Final states & Exchanged particles \\
\Xhline{0.8pt}
				$D^*K^*$ & $\pi$\\
\Xhline{0.4pt}
				$DK$, $D^*K$, $DK^*$ $D_1K$ & $\rho$, $\omega$\\
\Xhline{0.4pt}
				$D_s^*\eta$, $D^*_s\eta'$, $D_s\eta$, $D_s\eta'$ & $K^*$, \underline{$D^*$}\\

\Xhline{0.4pt}
				$D_s^*\omega$ & $K$, \underline{$K^*$}, \underline{$D$}, {\underline{$D^*$}}\\
\Xhline{0.4pt}
				$D_s^*\phi$ & $K$, $\underline{K^*}$\\
				\Xhline{1.0pt}
			\end{tabular}
\end{spacing}
	\end{threeparttable}
\end{table}

The coupling between a molecule and its components can be estimated model-independently via Weinberg compositeness criterion~\cite{Weinberg:1965zz,Baru:2003qq}, namely,
\begin{align}
\mathcal L_{XDK_1}=y X^{\mu}D^\dagger K^\dagger_{1\mu}+ h.c.
\end{align}
with
\begin{equation}
y^2=16\pi(m_{D}+m_{K_1})^{5/2}\sqrt{\frac{2E_b}{m_{D}m_{K_1}}}.\label{eq:Wein}
\end{equation}
In the following the binding energy will be fixed to $E_b=20$ MeV since Eq.(\ref{eq:Wein}) mainly contributes an overall factor to the total decay width and has little effect on the branching ratios of different channels.

To describe the transition from $DK_1$ to final states, the following Lagrangians, besides the ones introduced in the previous section, are needed.
\begin{align}
\mathcal{L}_{APV}&= i a\ {\rm Tr}(A_\mu[V^\mu,P]),\\
\mathcal{L}_{BPV}&= b\ {\rm{Tr}}(B_\mu\{V^\mu,P\})\\
\mathcal{L}_{AVV}&= a'\epsilon^{\alpha\beta\gamma\delta} \left(\partial_\alpha K_{1, A \beta}^{+}\rho^{-}_\gamma \bar{K}^{*0}_\delta\right.\notag\\
&\left.+\partial_\alpha K_{1, A \beta}^{-}\rho^{+}_\gamma {K}^{*0}_\delta+\cdots\right)\\
\mathcal{L}_{BVV}&=ib'\epsilon^{\alpha\beta\gamma\delta}\left(\partial_\alpha K_{1, B \beta}^{+}\rho^{-}_\gamma \bar{K}^{*0}_\delta\right.\notag\\
&\left.-\partial_\alpha K_{1, B \beta}^{-}\rho^{+}_\gamma {K}^{*0}_\delta+\cdots\right)\\
\mathcal{L}_{{DD}^{*} {P}}&=-\frac{2 g}{f_{\pi}} \sqrt{{m}_{{D}}{m}_{{D}^{*}}}\left(D_b\left(\partial_{\mu} P_{b a}\right) {D}_{a}^{* \mu \dagger}\right.\notag\\
&\left.- \overline{{D}}_{a}^{* \mu \dagger}\left(\partial_{\mu} P_{a b}\right) \overline{{D}}_{b}\right)\\
\mathcal{L}_{D^{*} {DV}}&=i\sqrt{2} \lambda g_{V}\epsilon_{\alpha \beta \gamma \delta}\left((\partial^\delta D_{a}) (\partial^{\alpha}V^{\beta}_{{ab}}) D^{* \dagger \gamma}_b\right.\notag\\
&\left.- D_{a} (\partial^{\alpha}V^{\beta}_{{ab}}) (\partial^\delta D^{* \dagger \gamma}_b)\right)+h.c.
\end{align}
where the coupling constants are $g=0.3 \sim 0.6$~\cite{Casalbuoni:1996pg}, $f_\pi=132$ MeV, $\lambda=0.6\ {\rm GeV}^{-1}$~\cite{Casalbuoni:1992dx}. In Ref.~\cite{Divotgey:2013jba} the coupling constants corresponding to the mixing angle $\phi=(56.4\pm4.3)^{\circ}$ are estimated to be $a\approx1.92\pm0.09$ GeV and $b\approx-2.47\pm0.08$ GeV. The coupling constants after mixing are listed in Tab.(\ref{mixedcoupling}). Unlike $a$ and $b$ determined by partial decay widths of $K_1$, no direct experimental results are available for $a'$ and $b'$ and we use a quark model approach to estimate them as presented in the Appendix.

\begin{table*}[htpb]
	\centering
	\caption{\label{mixedcoupling}Coupling constants of $K_1(1400)/K_1(1270)PV$.}
\begin{threeparttable}
\begin{spacing}{1.4}
\begin{tabular}{c|c|c}
\Xhline{1.0pt}
	 & $K_1(1400)$&$K_1(1270)$ \\
\Xhline{0.8pt}
\ \ $K\rho$\ \ & \ \ $-a \sin(\phi)-b\cos(\phi)$\ \ &\ \ $a\cos(\phi)-b\sin(\phi)$\ \ \\
\Xhline{0.8pt}
\ \ $K\omega$\ \ & \ \ $\sqrt{1/2}(-a \sin(\phi)-b\cos(\phi))$\ \ &\ \ $\sqrt{1/2}(a\cos(\phi)-b\sin(\phi))$\ \ \\
\Xhline{0.8pt}
\ \ $K\phi$\ \ & \ \ $a \sin(\phi)-b\cos(\phi)$\ \ &\ \ $a\cos(\phi)+b\sin(\phi)$\ \ \\
\Xhline{0.8pt}
\ \ $K^*\pi$\ \ & \ \ $a \sin(\phi)-b\cos(\phi)$\ \ &\ \ $a\cos(\phi)+b\sin(\phi)$\ \ \\
\Xhline{0.8pt}
\ \ $K^*\eta_8$\ \ & \ \ $\sqrt{3/2}\,a \sin(\phi)+\sqrt{1/6}\, b\cos(\phi)$\ \ &\ \ $-\sqrt{3/2}\, a \cos(\phi)+\sqrt{1/6}\, b\sin(\phi)$\ \ \\
\Xhline{0.8pt}
\ \ $K^*\eta_0$\ \ & \ \ $-\sqrt{4/3}\, b \cos(\phi)$\ \ &\ \ $-\sqrt{4/3}\, b\sin(\phi)$\ \ \\
\Xhline{1.0pt}			
\end{tabular}
\end{spacing}
	\end{threeparttable}
\end{table*}

When performing the loop integral in the triangle diagram of two-body decays, a Gaussian form factor 
\begin{equation}
F_0(p,\Lambda_0)=e^{-\bm{p}^2/\Lambda_0^2}
\end{equation}
and a monopole form factor, Eq.(\ref{eq:form_factor}), are introduced to the $XDK_1$ vertex and the propagator of the exchanged particle $A$, respectively. $\bm{p}$ is the three-momentum of the components in the rest frame of the molecule and $\Lambda_0$ plays the role of cutting high momentum components off in the wave function of the molecule.

\begin{table*}
\centering
\caption{\label{tab:width1400}Partial decay width of the predicted $DK_1(1400)$ molecule. Here cutoffs are in unit of GeV while widths are in unity of MeV. $\alpha$E$\beta$ means $\alpha\times10^{\beta}$. For the $DK^*$ channel, the sign of $a'b'$ is not determined and the results for both cases are listed here.}
\begin{threeparttable}
\begin{spacing}{1.4}
\begin{tabular}{|c|c|c|c|c|c|c|}
\Xhline{1.0pt}
\multirow{2}{*}{ Final states} &\multicolumn{3}{c|}{\ \ $\Lambda_0=0.6$\ \ }  &\multicolumn{3}{c|}{\ \ $\Lambda_0=1.0$\ \ } \\ \cline{2-7} 
& \ \ $\Lambda=1.5$\ \          & \ \ $\Lambda=2.0$   \ \      &\ \  $\Lambda=100$ \ \    &\ \ $\Lambda=1.5$\ \          & \ \ $\Lambda=2.0$   \ \      &\ \  $\Lambda=100$ \ \ \\
\Xhline{0.8pt}
 $D^*K^*$ & 3.15 & 4.42 & 7.70 & 6.09 & 9.06 & 18.8 \\
\Xhline{0.4pt}
 $D^*K$ & 7.1E-4 & 2.1E-3 & 0.010 & 1.9E-3 & 6.2E-3 & 0.035 \\
\Xhline{0.4pt}
 $DK^*\ (a'b'>0)$ & 7.1E-3 & 0.012 & 0.11 & 0.020 & 0.068 & 0.48 \\
  \Xhline{0.4pt}
 $DK^*\ (a'b'<0)$ & 3.7E-6 & 6.0E-6 & 5.7E-5 & 1.0E-5 & 3.5E-5 & 2.5E-4 \\
\Xhline{0.4pt}
 $DK$ & 1.4E-4 & 3.4E-4 & 7.8E-4 & 9.5E-4 & 2.2E-3 & 3.3E-3 \\
\Xhline{0.4pt}
 $D_s\eta$  & 8.6E-4 & 2.2E-3 & 4.1E-3 & 6.3E-3 & 0.017 & 0.022 \\
\Xhline{0.4pt}
 $D_s\eta '$ & 9.7E-4 & 2.3E-3 & 3.7E-3 & 5.9E-3 & 0.015 & 0.019 \\
\Xhline{0.4pt}
 $D_s^*\eta$  & 6.8E-3 & 0.025 & 0.12 & 0.018 & 0.069 & 0.44 \\
\Xhline{0.4pt}
 $D_s^*\eta'$ & 5.7E-3 & 0.020 & 0.096 & 0.014 & 0.055 & 0.33 \\
\Xhline{0.4pt}
 $D_1K$ & 3.1E-3 & 6.0E-3 & 0.013 & 8.5E-3 & 0.018 & 0.048 \\
\Xhline{0.4pt}
 $D_s^*\omega$  & 2.9E-3 & 4.9E-3 & 0.011 & 6.2E-3 & 0.011 & 0.030 \\
\Xhline{0.4pt}
 $D_s^*\phi$  & 0.47 & 0.73 & 1.35 & 0.83 & 1.38 & 3.11 \\
\Xhline{0.4pt}
$DK\rho$ &  \multicolumn{6}{c|}{\ \ $0.049$\ \ }  \\
\Xhline{0.4pt}
$DK\omega$ &  \multicolumn{6}{c|}{\ \ $0.015$\ \ } \\
\Xhline{0.4pt}
$DK^*\pi$ &  \multicolumn{6}{c|}{\ \ $27.0$\ \ } \\
\Xhline{0.6pt}
Total &31 &32& 36 &34& 38 &50\\
\Xhline{1.0pt}
\end{tabular}
\end{spacing}
\end{threeparttable}
\end{table*}

\begin{table*}[htpb]
	\centering
	\caption{\label{tab:width1270}Partial decay width of the predicted $DK_1(1270)$ molecule. Same captions as Tab.(\ref{tab:width1400})}
	\begin{threeparttable}
\begin{spacing}{1.4}
			\begin{tabular}{|c|c|c|c|c|c|c|}
				\Xhline{1.0pt}
				\multirow{2}{*}{ Final states} &\multicolumn{3}{c|}{\ \ $\Lambda_0=0.6$\ \ }  &\multicolumn{3}{c|}{\ \ $\Lambda_0=1.0$\ \ } \\ \cline{2-7} 
                         & \ \ $\Lambda=1.5$\ \          & \ \ $\Lambda=2.0$   \ \      &\ \  $\Lambda=100$ \ \        &\ \ $\Lambda=1.5$\ \          & \ \ $\Lambda=2.0$   \ \      &\ \  $\Lambda=100$ \ \        \\
				\Xhline{0.8pt}
 $D^*K^*$ & 0.63& 0.81 & 1.22 & 1.00 & 1.37 & 2.38 \\
\Xhline{0.4pt}
 $D^*K$ & 0.22 & 0.62 & 2.50 & 0.53 & 1.58 & 7.74 \\
\Xhline{0.4pt}
 $DK^*\ (a'b'>0)$ & 1.5E-3 & 1.5E-3 & 0.024 & 2.6E-3 & 0.022 & 0.090 \\
  \Xhline{0.4pt}
 $DK^*\ (a'b'<0)$ & 9.0E-3 & 9.4E-3 & 0.016 & 0.015 & 0.13 & 0.54 \\
\Xhline{0.4pt}
 $DK$ & 0.047 & 0.11 & 0.22 & 0.26 & 0.58 & 0.87 \\
\Xhline{0.4pt}
 $D_s\eta$  & 3.4E-3 & 8.5E-3 & 0.014 & 0.021 & 0.055 & 0.072 \\
\Xhline{0.4pt}
 $D_s\eta '$ &2.6E-3 & 5.7E-3 & 8.7E-3 & 0.013 & 0.031 & 0.040 \\
\Xhline{0.4pt}
 $D_s^*\eta$  & 0.025 & 0.084 & 0.37 & 0.056 & 0.21 & 1.14 \\
\Xhline{0.4pt}
 $D_s^*\eta'$ & 3.3E-3 & 9.7E-3 & 0.033 & 6.6E-3 & 0.021 &
   0.092 \\
\Xhline{0.4pt}
 $D_1K$ &  0.96 & 1.72 & 3.43 & 2.25 & 4.39 & 10.9 \\
\Xhline{0.4pt}
 $D_s^*\omega$  &  0.77 & 1.20 & 2.30 & 1.37 & 2.31 & 5.26 \\
\Xhline{0.4pt}
$DK\rho\to DK\pi\pi$ &  \multicolumn{6}{c|}{\ \ $4.56$ \ \ }  \\
\Xhline{0.4pt}
$DK^*\pi$ &  \multicolumn{6}{c|}{\ \ $3.32$ \ \ } \\
\Xhline{0.4pt}
$D\pi K_0^*(1430)\to DK\pi\pi$ &  \multicolumn{6}{c|}{\ \ $5.69$ \ \ }  \\
\Xhline{0.6pt}
Total &16 &18& 24 &19& 24 &42\\
\Xhline{1.0pt}
\end{tabular}
\end{spacing}
	\end{threeparttable}
\end{table*}

The results for partial widths of the molecules are listed in Table~\ref{tab:width1400} and Table~\ref{tab:width1270}. In our model, the absolute values of widths may suffer some uncertainty, resulting from the uncertainty of coupling constants as well as the choice of cutoffs. The former just mainly gives a scaling factor of order one to each channel.  The latter one may also change the branching ratios of different channels but it turns out that the dominance of some decay channels is not influenced by the cutoffs. Due to the mixture nature of $K_1(1400)$ and $K_1(1270)$, the couplings of $K_1K\rho/\omega$ and $K_1K^*\pi$ are quite different. For the $DK_1(1400)$ molecule, three-body channel $DK^*\pi$ dominates and $D^*K^*$ channel also has considerable contribution due to the $\pi$ exchange. While for the $DK_1(1270)$ molecule, the partial decay widths to $D^*K$, $DK$, $D_1K$, $D_s^*\eta$ and $D_s^*\omega$ channels are not that small as the $K_1(1400)$ case because of the much larger coupling of $K_1(1270)K\rho/\omega$. Meanwhile the width of three-body channels become much smaller because of either small coupling of $K_1(1270)K\pi$ or vanishing phase space for $DK\rho/\omega$ and $DK_0^*(1430)\pi$. The finite width of $\rho$ or $K_0^*(1430)$ needs to be considered because $DK_1(1270)$ molecule lies below the threshold of $DK\rho$ or $DK_0^*(1430)\pi$ but $\rho$ and $K_0^*(1430)$ have quite large width. Here we consider the four body decay, $DK_1(1270)\to DK\rho (DK_0^*\pi)\to DK\pi\pi$ and refer to Ref.~\cite{Jing:2020tth} for elegant calculations of phase space integration.

\section{Summary and Discussion}
In this work we have calculate the potential between $DK_1$ or $D_1K$ systems to check if they are  possible to form bound states. The interaction between these components are described by vector meson exchange model with effective Lagrangians. It turns out that for isoscalar systems, $DK_1$ or $D_1K$ are attractive but only in
the  former case is it possible to form bound states if the cutoff lies in the empirical region. Isovector systems have either repulsive or too weak attractive potentials and therefore, no bound states are expected. The $X_1(2900)$, recently reported by LHCb collaboration, makes it meaningful to explore the possibility of $\bar D_1 K$ bound states since they can couple in S-wave and $X_1(2900)$ lies just about 10 MeV below  $\bar D_1 K$ threshold. Our results show that $\bar D_1 K$ system, no matter isoscalar or isovector, can not be bound together via one meson exchange and the explanation of $X_1(2900)$ as a $\bar D_1 K$ is disfavored. 

The decay properties of the predicted isoscalar $DK_1$ molecules are calculated by applying Weinberg compositeness criterion. For the $DK_1(1400)$ bound state, due to the large width of $K_1(1400)\to K^*\pi$, the three-body channel $DK^*\pi$ dominates the total width of the bound states. Besides, $D^*K^*$ and $D_s^*\phi$ channels are also good places to search for the predicted molecule. While for the $DK_1(1270)$ bound state, it may be rewarding to look at the $D^*K^*$, $D^*K$, $D_1K$, $D_s^*\omega$ and $DK^*\pi$ channels.

The heavy quark symmetry allows one to predicted the corresponding bound states of $BK_1$ systems, whose binding energy are around $50$ MeV and the decay behaviors should be similar to those of $DK_1$ bound states.

Helpful discussions with Feng-Kun Guo, Hao-Jie Jing, Yong-Hui Lin and Mao-Jun Yan are acknowledged. This project is supported by NSFC under Grant No. 11621131001 (CRC110 cofunded by DFG and NSFC), Grant No. 11835015,  No. 11947302, and by the Chinese Academy of Sciences (CAS) under Grants No. XDB34030303.

\appendix
\section{Estimation of $g_{K_1 K^* V}$}
\subsection{Effective Lagrangian on Hadron Level}
The coupling of $K_1KV$ on hadron level can be constructed as
\begin{align}
\mathcal{L}_{APV}&=i a \operatorname{Tr}\left(A_{\mu}\left[V^{\mu}, P\right]\right)
=-ia K_{1, A \mu}^{+}\rho^{\mu-} \bar{K}^{0}+\cdots\\
\mathcal{L}_{BPV}&=b \operatorname{Tr}\left(B_{\mu}\left\{V^{\mu}, P\right\}\right)
=bK_{1, B \mu}^{+}\rho^{\mu-} \bar{K}^{0}+\cdots
\end{align}
while for $K_1K^*V$,
\begin{align}
\mathcal{L}_{AVV}
=& a'\epsilon^{\alpha\beta\gamma\delta} \left(\partial_\alpha K_{1, A \beta}^{+}\rho^{-}_\gamma \bar{K}^{*0}_\delta\right.\notag\\
&\left.+\partial_\alpha K_{1, A \beta}^{-}\rho^{+}_\gamma {K}^{*0}_\delta+\cdots\right)\\
\mathcal{L}_{BVV}
=&ib'\epsilon^{\alpha\beta\gamma\delta}\left(\partial_\alpha K_{1, B \beta}^{+}\rho^{-}_\gamma \bar{K}^{*0}_\delta\right.\notag\\
&\left.-\partial_\alpha K_{1, B \beta}^{-}\rho^{+}_\gamma {K}^{*0}_\delta+\cdots\right)..
\end{align}
We now aim at estimating $a',b'$ from $a,b$. Considering the mixing, Eq.(\ref{eq:mixing}), we have 
\begin{align}
g_{K_1(1270)K^*\rho}=&(a'\cos\varphi+b'\sin(\varphi))\\
g_{K_1(1400)K^*\rho}=&i(b'\cos\varphi-a'\sin(\varphi)).
\end{align}

In the nonrelativistic (NR) limit, the squared amplitudes after sum of polarizations read
\begin{align}
\frac13\sum|\mathcal M|^2
\approx\left\{\begin{array}{ll}
    (a\ {\rm or\ } b)^2 & \mathrm{for}\  K_1\to K\rho \\
   (a'\ {\rm or }\ b')^2 m_{K_1}^2 &\mathrm{for}\ K_1\to K^*\rho 
\end{array}\right.
\end{align}

\subsection{Effective Lagrangian from Quark Model}
 Analogous to the quark model describing $\gamma N$ interaction, see Ref.~\cite{Close:1979} for a detailed discussion, the effective coupling of $K_1K^{(*)}V$ (essentially $qqV$ and we take $\rho$ meson as an example) can be constructed as
\begin{equation}
\mathcal{L}=g\left[2i\bm{s}^j\cdot(\bm k\times \bm{\rho})+2\bm{p}\cdot\bm{\rho}\right]
\end{equation}
with $g$ the coupling constant, $\bm{s}$ the spin operator, $\bm{\rho}$ the vector field, $\bm k$ the 3-momentum of the vector and $\bm q$ the 3-momentum of the quark. This Lagrangian can be derived from the quark level. For a free fermion the Hamiltonian reads 
\begin{equation}
H_{\rm{free}}=\frac{p^2}{2m}=\frac{\slashed{p}\slashed{p}}{2m}.
\end{equation}
We introduced the coupling of a vector field to the quark by the following remedy,
 \begin{equation}
H=\frac{(\slashed{p}-g\slashed{\rho})(\slashed{p}-g\slashed{\rho})}{2m}=H_{\rm{free}}+H_{\rm{int}}
\end{equation}
where
\begin{align}
H_{\rm{int}}&\approx -g(\slashed{p}\slashed{\rho}+\slashed{\rho}\slashed{p})
\end{align}
with 
\begin{align}
\rho_\mu(x)=\int \frac{d^{3} k}{(2 \pi)^{3}} \frac{1}{\sqrt{2E_{\bm{k}}}}&\left(\epsilon_\mu(\bm{k})a_{\bm{k}} e^{-i p \cdot x}\right.\notag\\
&\left.+\left.\epsilon^*_\mu(\bm{k})a_{\mathbf{k}}^{\dagger} e^{i k \cdot x}\right)\right|_{k^{0}=E_{\mathbf{k}}}
\end{align}
After contracting with the external vector meson we have
\begin{align}
H_{\rm{int}}&\propto-g\frac{e^{ik\cdot x}}{\sqrt{2k^0}}(2p\cdot\epsilon^*+\slashed{k}\slashed{\epsilon}^*).
\end{align}

Considering the Dirac representation of $\gamma$ matrices
\begin{align}
\gamma^0&=\left(\begin{array}{cc}
1 & 0 \\
0 & -1
\end{array}\right),\\
\gamma^i&=\left(\begin{array}{cc}
0 & \sigma^i \\
-\sigma^i & 0
\end{array}\right),\notag\\
\alpha^i&=\gamma^0\gamma^i=\left(\begin{array}{cc}
0 & \sigma^i \\
\sigma^i & 0
\end{array}\right),
\end{align}
we obtain
\begin{align}
H_{\rm{int}}&\propto g\frac{e^{ik\cdot x}}{\sqrt{2k^0}}\left[2\bm{p}\cdot\bm{\epsilon}^*+i\bm{\sigma}\cdot(\bm{k}\times\bm{\epsilon}^*)\right.\notag\\
&\left.-2(p^0\epsilon^{*0}+k^0\epsilon^{*0})+(k^0\bm{\epsilon}^*-\epsilon^{*0}\bm{k})\cdot\bm{\alpha}\right]
\end{align}
In the NR limit, $\epsilon^0=0$, $k^0=m_\rho$ and $\bm{\epsilon}^*\cdot\bm{\alpha}$ mix the large and small components of the spinor. After omitting these terms we obtain
\begin{align}
H_{\rm{int}}&\propto g\frac{e^{ik\cdot x}}{\sqrt{2m_\rho}}\left[2\bm{p}\cdot\bm{\epsilon}^*+i\bm{\sigma}\cdot(\bm{k}\times\bm{\epsilon}^*)\right]\notag\\
&\propto e^{ik\cdot x}\left[2\bm{p}\cdot\bm{\epsilon}^*+i\bm{\sigma}\cdot(\bm{k}\times\bm{\epsilon}^*)\right].
\end{align}
Note that the first term flips $L_z$ while the second one flips $S_z$.

The wave functions of related particles are collected in the following
\begin{align}
\ket{K_{1A}(1,m)}=&\sum_{m_1+m_2=m}\bracket{1,m_1,1,m_2}{1,m}\Psi_{1,m_1}\chi_{1,m_2}\\
\ket{K_{1B}(1,m)}=&\Psi_{1,m}\chi_{0,0}\\
\ket{\rho(1,m)}=&\chi_{1,m}\\
\ket{K}=&\Psi_{0,0}\chi_{0,0}\\
\ket{K^*(1,m)}=&\Psi_{0,0}\chi_{1,m}
\end{align}
where $\Psi$ and $\chi$ represent the spatial and spin wave functions, respectively. Then let's fix the kinetics. $K_{1A/B}$ decays into $K^{(*)}\rho$ where $K_{1}$ is at rest with $P=(m_{K_{1}},0,0,0)$ and $\rho$ flies along the $z-$axis with $k=(m_\rho,0,0,k)$. The polarization of $\rho$, represented by the $\bm{\epsilon}$ in the Lagrangian, reads explicitly
\begin{align}
\bm{\epsilon}(1)&=-\frac{1}{\sqrt{2}}(1,i,0)\\
\bm{\epsilon}(0)&=(0,0,1)\\
\bm{\epsilon}(-1)&=\frac{1}{\sqrt{2}}(1,-i,0)
\end{align}
and in turn
\begin{align}
2\bm{p}\cdot\bm{\epsilon}^*+i\bm{\sigma}\cdot(\bm{k}\times\bm{\epsilon}^*)=\left\{\begin{array}{ll}
{\sqrt2}p_--\frac{k}{\sqrt2}\sigma_-&\ {\rm for}\ \bm{\epsilon}(1)\\
2p_z&\ {\rm for}\ \bm{\epsilon}(0)\\
{\sqrt2}p_++\frac{k}{\sqrt2}\sigma_+&\ {\rm for}\ \bm{\epsilon}(-1)\\
\end{array}\right.
\end{align}
with $\sigma_{\pm}=\sigma_x\pm i\sigma_y$ and $p_{\pm}=p_x\pm ip_y$.

Now the calculation of the amplitude of $K_{1A/B}\to K^{(*)}\rho$ is straightforward. Let's first define the following constants
\begin{align}
l_-=&\left|\bra{\Psi_{0,0}}e^{-ikz}\sqrt2 p_-\ket{\Psi_{1,1}}\right|^2\\
l_+=&\left|\bra{\Psi_{0,0}}e^{-ikz}\sqrt2 p_+\ket{\Psi_{1,-1}}\right|^2\\
l_0=&\left|\bra{\Psi_{0,0}}e^{-ikz}2 p_z\ket{\Psi_{1,0}}\right|^2\\
s_0=&\left|\frac{k}{2\sqrt2}\bra{\Psi_{0,0}}e^{-ikz}\ket{\Psi_{1,0}}\right|^2
\end{align}
and then we have
\begin{widetext}
\begin{align}
\frac13\sum |\mathcal M|^2=\left\{
\begin{array}{lll}
\frac23 s_0 &=a^2&\ {\rm for}\ K_{1A}\to K\rho\\
\frac13 (l_++l_-+l_0)&=b^2&\ {\rm for}\ K_{1B}\to K\rho\\
\frac13\left( s_0+\frac12(l_++l_-+l_0)\right)&=2a^{'2}m_{K_{1A}}^2&\ {\rm for}\ K_{1A}\to K^*\rho\\
\frac13 s_0&=2b^{'2}m_{K_{1B}}^2&\ {\rm for}\ K_{1B}\to K^*\rho
\end{array}\right.
\end{align}
\end{widetext}

$a=1.92\ {\rm GeV}$, $b=-2.47\ {\rm GeV}$, $m_{K_{1 A}}=1.36\ \mathrm{GeV}$ and $m_{K_{1 B}}=1.31\ \mathrm{GeV}$ as inputs yield $|a'|=1.15$ and $|b'|=0.73$. The relative sign of $a'$ and $b'$ can not be determined within this model.

\bibliography{ref}

\begin{thebibliography}{64}%
\makeatletter
\providecommand \@ifxundefined [1]{%
 \@ifx{#1\undefined}
}%
\providecommand \@ifnum [1]{%
 \ifnum #1\expandafter \@firstoftwo
 \else \expandafter \@secondoftwo
 \fi
}%
\providecommand \@ifx [1]{%
 \ifx #1\expandafter \@firstoftwo
 \else \expandafter \@secondoftwo
 \fi
}%
\providecommand \natexlab [1]{#1}%
\providecommand \enquote  [1]{``#1''}%
\providecommand \bibnamefont  [1]{#1}%
\providecommand \bibfnamefont [1]{#1}%
\providecommand \citenamefont [1]{#1}%
\providecommand \href@noop [0]{\@secondoftwo}%
\providecommand \href [0]{\begingroup \@sanitize@url \@href}%
\providecommand \@href[1]{\@@startlink{#1}\@@href}%
\providecommand \@@href[1]{\endgroup#1\@@endlink}%
\providecommand \@sanitize@url [0]{\catcode `\\12\catcode `\$12\catcode
  `\&12\catcode `\#12\catcode `\^12\catcode `\_12\catcode `\%12\relax}%
\providecommand \@@startlink[1]{}%
\providecommand \@@endlink[0]{}%
\providecommand \url  [0]{\begingroup\@sanitize@url \@url }%
\providecommand \@url [1]{\endgroup\@href {#1}{\urlprefix }}%
\providecommand \urlprefix  [0]{URL }%
\providecommand \Eprint [0]{\href }%
\providecommand \doibase [0]{http://dx.doi.org/}%
\providecommand \selectlanguage [0]{\@gobble}%
\providecommand \bibinfo  [0]{\@secondoftwo}%
\providecommand \bibfield  [0]{\@secondoftwo}%
\providecommand \translation [1]{[#1]}%
\providecommand \BibitemOpen [0]{}%
\providecommand \bibitemStop [0]{}%
\providecommand \bibitemNoStop [0]{.\EOS\space}%
\providecommand \EOS [0]{\spacefactor3000\relax}%
\providecommand \BibitemShut  [1]{\csname bibitem#1\endcsname}%
\let\auto@bib@innerbib\@empty
\bibitem [{\citenamefont {Aaij}\ \emph {et~al.}(2020)\citenamefont {Aaij} \emph
  {et~al.}}]{Aaij:2020ypa}%
  \BibitemOpen
  \bibfield  {author} {\bibinfo {author} {\bibfnamefont {R.}~\bibnamefont
  {Aaij}} \emph {et~al.} (\bibinfo {collaboration} {LHCb}),\ }\href {\doibase
  10.1103/PhysRevD.102.112003} {\bibfield  {journal} {\bibinfo  {journal}
  {Phys. Rev. D}\ }\textbf {\bibinfo {volume} {102}},\ \bibinfo {pages}
  {112003} (\bibinfo {year} {2020})},\ \Eprint
  {http://arxiv.org/abs/2009.00026} {arXiv:2009.00026 [hep-ex]} \BibitemShut
  {NoStop}%
\bibitem [{\citenamefont {Karliner}\ and\ \citenamefont
  {Rosner}(2020)}]{Karliner:2020vsi}%
  \BibitemOpen
  \bibfield  {author} {\bibinfo {author} {\bibfnamefont {M.}~\bibnamefont
  {Karliner}}\ and\ \bibinfo {author} {\bibfnamefont {J.~L.}\ \bibnamefont
  {Rosner}},\ }\href {\doibase 10.1103/PhysRevD.102.094016} {\bibfield
  {journal} {\bibinfo  {journal} {Phys. Rev. D}\ }\textbf {\bibinfo {volume}
  {102}},\ \bibinfo {pages} {094016} (\bibinfo {year} {2020})},\ \Eprint
  {http://arxiv.org/abs/2008.05993} {arXiv:2008.05993 [hep-ph]} \BibitemShut
  {NoStop}%
\bibitem [{\citenamefont {He}\ \emph {et~al.}(2020)\citenamefont {He},
  \citenamefont {Wang},\ and\ \citenamefont {Zhu}}]{He:2020jna}%
  \BibitemOpen
  \bibfield  {author} {\bibinfo {author} {\bibfnamefont {X.-G.}\ \bibnamefont
  {He}}, \bibinfo {author} {\bibfnamefont {W.}~\bibnamefont {Wang}}, \ and\
  \bibinfo {author} {\bibfnamefont {R.}~\bibnamefont {Zhu}},\ }\href {\doibase
  10.1140/epjc/s10052-020-08597-1} {\bibfield  {journal} {\bibinfo  {journal}
  {Eur. Phys. J. C}\ }\textbf {\bibinfo {volume} {80}},\ \bibinfo {pages}
  {1026} (\bibinfo {year} {2020})},\ \Eprint {http://arxiv.org/abs/2008.07145}
  {arXiv:2008.07145 [hep-ph]} \BibitemShut {NoStop}%
\bibitem [{\citenamefont {Zhang}(2021)}]{Zhang:2020oze}%
  \BibitemOpen
  \bibfield  {author} {\bibinfo {author} {\bibfnamefont {J.-R.}\ \bibnamefont
  {Zhang}},\ }\href {\doibase 10.1103/PhysRevD.103.054019} {\bibfield
  {journal} {\bibinfo  {journal} {Phys. Rev. D}\ }\textbf {\bibinfo {volume}
  {103}},\ \bibinfo {pages} {054019} (\bibinfo {year} {2021})},\ \Eprint
  {http://arxiv.org/abs/2008.07295} {arXiv:2008.07295 [hep-ph]} \BibitemShut
  {NoStop}%
\bibitem [{\citenamefont {Wang}(2020)}]{Wang:2020xyc}%
  \BibitemOpen
  \bibfield  {author} {\bibinfo {author} {\bibfnamefont {Z.-G.}\ \bibnamefont
  {Wang}},\ }\href {\doibase 10.1142/S0217751X20501870} {\bibfield  {journal}
  {\bibinfo  {journal} {Int. J. Mod. Phys. A}\ }\textbf {\bibinfo {volume}
  {35}},\ \bibinfo {pages} {2050187} (\bibinfo {year} {2020})},\ \Eprint
  {http://arxiv.org/abs/2008.07833} {arXiv:2008.07833 [hep-ph]} \BibitemShut
  {NoStop}%
\bibitem [{\citenamefont {L\"u}\ \emph {et~al.}(2020)\citenamefont {L\"u},
  \citenamefont {Chen},\ and\ \citenamefont {Dong}}]{Lu:2020qmp}%
  \BibitemOpen
  \bibfield  {author} {\bibinfo {author} {\bibfnamefont {Q.-F.}\ \bibnamefont
  {L\"u}}, \bibinfo {author} {\bibfnamefont {D.-Y.}\ \bibnamefont {Chen}}, \
  and\ \bibinfo {author} {\bibfnamefont {Y.-B.}\ \bibnamefont {Dong}},\ }\href
  {\doibase 10.1103/PhysRevD.102.074021} {\bibfield  {journal} {\bibinfo
  {journal} {Phys. Rev. D}\ }\textbf {\bibinfo {volume} {102}},\ \bibinfo
  {pages} {074021} (\bibinfo {year} {2020})},\ \Eprint
  {http://arxiv.org/abs/2008.07340} {arXiv:2008.07340 [hep-ph]} \BibitemShut
  {NoStop}%
\bibitem [{\citenamefont {Chen}\ \emph {et~al.}(2020)\citenamefont {Chen},
  \citenamefont {Chen}, \citenamefont {Dong},\ and\ \citenamefont
  {Su}}]{Chen:2020aos}%
  \BibitemOpen
  \bibfield  {author} {\bibinfo {author} {\bibfnamefont {H.-X.}\ \bibnamefont
  {Chen}}, \bibinfo {author} {\bibfnamefont {W.}~\bibnamefont {Chen}}, \bibinfo
  {author} {\bibfnamefont {R.-R.}\ \bibnamefont {Dong}}, \ and\ \bibinfo
  {author} {\bibfnamefont {N.}~\bibnamefont {Su}},\ }\href {\doibase
  10.1088/0256-307X/37/10/101201} {\bibfield  {journal} {\bibinfo  {journal}
  {Chin. Phys. Lett.}\ }\textbf {\bibinfo {volume} {37}},\ \bibinfo {pages}
  {101201} (\bibinfo {year} {2020})},\ \Eprint
  {http://arxiv.org/abs/2008.07516} {arXiv:2008.07516 [hep-ph]} \BibitemShut
  {NoStop}%
\bibitem [{\citenamefont {Hu}\ \emph {et~al.}(2021)\citenamefont {Hu},
  \citenamefont {Lao}, \citenamefont {Ling},\ and\ \citenamefont
  {Wang}}]{Hu:2020mxp}%
  \BibitemOpen
  \bibfield  {author} {\bibinfo {author} {\bibfnamefont {M.-W.}\ \bibnamefont
  {Hu}}, \bibinfo {author} {\bibfnamefont {X.-Y.}\ \bibnamefont {Lao}},
  \bibinfo {author} {\bibfnamefont {P.}~\bibnamefont {Ling}}, \ and\ \bibinfo
  {author} {\bibfnamefont {Q.}~\bibnamefont {Wang}},\ }\href {\doibase
  10.1088/1674-1137/abcfaa} {\bibfield  {journal} {\bibinfo  {journal} {Chin.
  Phys. C}\ }\textbf {\bibinfo {volume} {45}},\ \bibinfo {pages} {021003}
  (\bibinfo {year} {2021})},\ \Eprint {http://arxiv.org/abs/2008.06894}
  {arXiv:2008.06894 [hep-ph]} \BibitemShut {NoStop}%
\bibitem [{\citenamefont {Liu}\ \emph {et~al.}(2020{\natexlab{a}})\citenamefont
  {Liu}, \citenamefont {Xie},\ and\ \citenamefont {Geng}}]{Liu:2020nil}%
  \BibitemOpen
  \bibfield  {author} {\bibinfo {author} {\bibfnamefont {M.-Z.}\ \bibnamefont
  {Liu}}, \bibinfo {author} {\bibfnamefont {J.-J.}\ \bibnamefont {Xie}}, \ and\
  \bibinfo {author} {\bibfnamefont {L.-S.}\ \bibnamefont {Geng}},\ }\href
  {\doibase 10.1103/PhysRevD.102.091502} {\bibfield  {journal} {\bibinfo
  {journal} {Phys. Rev. D}\ }\textbf {\bibinfo {volume} {102}},\ \bibinfo
  {pages} {091502} (\bibinfo {year} {2020}{\natexlab{a}})},\ \Eprint
  {http://arxiv.org/abs/2008.07389} {arXiv:2008.07389 [hep-ph]} \BibitemShut
  {NoStop}%
\bibitem [{\citenamefont {He}\ and\ \citenamefont {Chen}(2020)}]{He:2020btl}%
  \BibitemOpen
  \bibfield  {author} {\bibinfo {author} {\bibfnamefont {J.}~\bibnamefont
  {He}}\ and\ \bibinfo {author} {\bibfnamefont {D.-Y.}\ \bibnamefont {Chen}},\
  }\href@noop {} {\  (\bibinfo {year} {2020})},\ \Eprint
  {http://arxiv.org/abs/2008.07782} {arXiv:2008.07782 [hep-ph]} \BibitemShut
  {NoStop}%
\bibitem [{\citenamefont {Huang}\ \emph {et~al.}(2020)\citenamefont {Huang},
  \citenamefont {Lu}, \citenamefont {Xie},\ and\ \citenamefont
  {Geng}}]{Huang:2020ptc}%
  \BibitemOpen
  \bibfield  {author} {\bibinfo {author} {\bibfnamefont {Y.}~\bibnamefont
  {Huang}}, \bibinfo {author} {\bibfnamefont {J.-X.}\ \bibnamefont {Lu}},
  \bibinfo {author} {\bibfnamefont {J.-J.}\ \bibnamefont {Xie}}, \ and\
  \bibinfo {author} {\bibfnamefont {L.-S.}\ \bibnamefont {Geng}},\ }\href
  {\doibase 10.1140/epjc/s10052-020-08516-4} {\bibfield  {journal} {\bibinfo
  {journal} {Eur. Phys. J. C}\ }\textbf {\bibinfo {volume} {80}},\ \bibinfo
  {pages} {973} (\bibinfo {year} {2020})},\ \Eprint
  {http://arxiv.org/abs/2008.07959} {arXiv:2008.07959 [hep-ph]} \BibitemShut
  {NoStop}%
\bibitem [{\citenamefont {Xue}\ \emph {et~al.}(2021)\citenamefont {Xue},
  \citenamefont {Jin}, \citenamefont {Huang},\ and\ \citenamefont
  {Ping}}]{Xue:2020vtq}%
  \BibitemOpen
  \bibfield  {author} {\bibinfo {author} {\bibfnamefont {Y.}~\bibnamefont
  {Xue}}, \bibinfo {author} {\bibfnamefont {X.}~\bibnamefont {Jin}}, \bibinfo
  {author} {\bibfnamefont {H.}~\bibnamefont {Huang}}, \ and\ \bibinfo {author}
  {\bibfnamefont {J.}~\bibnamefont {Ping}},\ }\href {\doibase
  10.1103/PhysRevD.103.054010} {\bibfield  {journal} {\bibinfo  {journal}
  {Phys. Rev. D}\ }\textbf {\bibinfo {volume} {103}},\ \bibinfo {pages}
  {054010} (\bibinfo {year} {2021})},\ \Eprint
  {http://arxiv.org/abs/2008.09516} {arXiv:2008.09516 [hep-ph]} \BibitemShut
  {NoStop}%
\bibitem [{\citenamefont {Molina}\ and\ \citenamefont
  {Oset}(2020)}]{Molina:2020hde}%
  \BibitemOpen
  \bibfield  {author} {\bibinfo {author} {\bibfnamefont {R.}~\bibnamefont
  {Molina}}\ and\ \bibinfo {author} {\bibfnamefont {E.}~\bibnamefont {Oset}},\
  }\href {\doibase 10.1016/j.physletb.2020.135870} {\bibfield  {journal}
  {\bibinfo  {journal} {Phys. Lett. B}\ }\textbf {\bibinfo {volume} {811}},\
  \bibinfo {pages} {135870} (\bibinfo {year} {2020})},\ \Eprint
  {http://arxiv.org/abs/2008.11171} {arXiv:2008.11171 [hep-ph]} \BibitemShut
  {NoStop}%
\bibitem [{\citenamefont {Agaev}\ \emph {et~al.}(2020)\citenamefont {Agaev},
  \citenamefont {Azizi},\ and\ \citenamefont {Sundu}}]{Agaev:2020nrc}%
  \BibitemOpen
  \bibfield  {author} {\bibinfo {author} {\bibfnamefont {S.}~\bibnamefont
  {Agaev}}, \bibinfo {author} {\bibfnamefont {K.}~\bibnamefont {Azizi}}, \ and\
  \bibinfo {author} {\bibfnamefont {H.}~\bibnamefont {Sundu}},\ }\href@noop {}
  {\  (\bibinfo {year} {2020})},\ \Eprint {http://arxiv.org/abs/2008.13027}
  {arXiv:2008.13027 [hep-ph]} \BibitemShut {NoStop}%
\bibitem [{\citenamefont {Liu}\ \emph {et~al.}(2020{\natexlab{b}})\citenamefont
  {Liu}, \citenamefont {Yan}, \citenamefont {Ke}, \citenamefont {Li},\ and\
  \citenamefont {Xie}}]{Liu:2020orv}%
  \BibitemOpen
  \bibfield  {author} {\bibinfo {author} {\bibfnamefont {X.-H.}\ \bibnamefont
  {Liu}}, \bibinfo {author} {\bibfnamefont {M.-J.}\ \bibnamefont {Yan}},
  \bibinfo {author} {\bibfnamefont {H.-W.}\ \bibnamefont {Ke}}, \bibinfo
  {author} {\bibfnamefont {G.}~\bibnamefont {Li}}, \ and\ \bibinfo {author}
  {\bibfnamefont {J.-J.}\ \bibnamefont {Xie}},\ }\href {\doibase
  10.1140/epjc/s10052-020-08762-6} {\bibfield  {journal} {\bibinfo  {journal}
  {Eur. Phys. J. C}\ }\textbf {\bibinfo {volume} {80}},\ \bibinfo {pages}
  {1178} (\bibinfo {year} {2020}{\natexlab{b}})},\ \Eprint
  {http://arxiv.org/abs/2008.07190} {arXiv:2008.07190 [hep-ph]} \BibitemShut
  {NoStop}%
\bibitem [{\citenamefont {Burns}\ and\ \citenamefont
  {Swanson}(2021{\natexlab{a}})}]{Burns:2020epm}%
  \BibitemOpen
  \bibfield  {author} {\bibinfo {author} {\bibfnamefont {T.~J.}\ \bibnamefont
  {Burns}}\ and\ \bibinfo {author} {\bibfnamefont {E.~S.}\ \bibnamefont
  {Swanson}},\ }\href {\doibase 10.1016/j.physletb.2020.136057} {\bibfield
  {journal} {\bibinfo  {journal} {Phys. Lett. B}\ }\textbf {\bibinfo {volume}
  {813}},\ \bibinfo {pages} {136057} (\bibinfo {year} {2021}{\natexlab{a}})},\
  \Eprint {http://arxiv.org/abs/2008.12838} {arXiv:2008.12838 [hep-ph]}
  \BibitemShut {NoStop}%
\bibitem [{\citenamefont {Albuquerque}\ \emph {et~al.}(2021)\citenamefont
  {Albuquerque}, \citenamefont {Narison}, \citenamefont {Rabetiarivony},\ and\
  \citenamefont {Randriamanatrika}}]{Albuquerque:2020ugi}%
  \BibitemOpen
  \bibfield  {author} {\bibinfo {author} {\bibfnamefont {R.~M.}\ \bibnamefont
  {Albuquerque}}, \bibinfo {author} {\bibfnamefont {S.}~\bibnamefont
  {Narison}}, \bibinfo {author} {\bibfnamefont {D.}~\bibnamefont
  {Rabetiarivony}}, \ and\ \bibinfo {author} {\bibfnamefont {G.}~\bibnamefont
  {Randriamanatrika}},\ }\href {\doibase 10.1016/j.nuclphysa.2020.122113}
  {\bibfield  {journal} {\bibinfo  {journal} {Nucl. Phys. A}\ }\textbf
  {\bibinfo {volume} {1007}},\ \bibinfo {pages} {122113} (\bibinfo {year}
  {2021})},\ \Eprint {http://arxiv.org/abs/2008.13463} {arXiv:2008.13463
  [hep-ph]} \BibitemShut {NoStop}%
\bibitem [{\citenamefont {Chen}\ \emph {et~al.}(2021)\citenamefont {Chen},
  \citenamefont {Han}, \citenamefont {L\"u}, \citenamefont {Wang},\ and\
  \citenamefont {Yu}}]{Chen:2020eyu}%
  \BibitemOpen
  \bibfield  {author} {\bibinfo {author} {\bibfnamefont {Y.-K.}\ \bibnamefont
  {Chen}}, \bibinfo {author} {\bibfnamefont {J.-J.}\ \bibnamefont {Han}},
  \bibinfo {author} {\bibfnamefont {Q.-F.}\ \bibnamefont {L\"u}}, \bibinfo
  {author} {\bibfnamefont {J.-P.}\ \bibnamefont {Wang}}, \ and\ \bibinfo
  {author} {\bibfnamefont {F.-S.}\ \bibnamefont {Yu}},\ }\href {\doibase
  10.1140/epjc/s10052-021-08857-8} {\bibfield  {journal} {\bibinfo  {journal}
  {J. Phys. G: Nucl. Part. Phys. 48 055007}\ }\textbf {\bibinfo {volume}
  {81}},\ \bibinfo {pages} {71} (\bibinfo {year} {2021})},\ \Eprint
  {http://arxiv.org/abs/2009.01182} {arXiv:2009.01182 [hep-ph]} \BibitemShut
  {NoStop}%
\bibitem [{\citenamefont {Mutuk}(2021)}]{Mutuk:2020igv}%
  \BibitemOpen
  \bibfield  {author} {\bibinfo {author} {\bibfnamefont {H.}~\bibnamefont
  {Mutuk}},\ }\href {\doibase 10.1088/1361-6471/abeb7f} {\bibfield  {journal}
  {\bibinfo  {journal} {J. Phys. G: Nucl. Part. Phys.}\ }\textbf {\bibinfo
  {volume} {48}},\ \bibinfo {pages} {055007} (\bibinfo {year} {2021})},\
  \Eprint {http://arxiv.org/abs/2009.02492} {arXiv:2009.02492 [hep-ph]}
  \BibitemShut {NoStop}%
\bibitem [{\citenamefont {Burns}\ and\ \citenamefont
  {Swanson}(2021{\natexlab{b}})}]{Burns:2020xne}%
  \BibitemOpen
  \bibfield  {author} {\bibinfo {author} {\bibfnamefont {T.~J.}\ \bibnamefont
  {Burns}}\ and\ \bibinfo {author} {\bibfnamefont {E.~S.}\ \bibnamefont
  {Swanson}},\ }\href {\doibase 10.1103/PhysRevD.103.014004} {\bibfield
  {journal} {\bibinfo  {journal} {Phys. Rev. D}\ }\textbf {\bibinfo {volume}
  {103}},\ \bibinfo {pages} {014004} (\bibinfo {year} {2021}{\natexlab{b}})},\
  \Eprint {http://arxiv.org/abs/2009.05352} {arXiv:2009.05352 [hep-ph]}
  \BibitemShut {NoStop}%
\bibitem [{\citenamefont {Wang}\ \emph {et~al.}(2021)\citenamefont {Wang},
  \citenamefont {Meng}, \citenamefont {Xiao}, \citenamefont {Oka},\ and\
  \citenamefont {Zhu}}]{Wang:2020prk}%
  \BibitemOpen
  \bibfield  {author} {\bibinfo {author} {\bibfnamefont {G.-J.}\ \bibnamefont
  {Wang}}, \bibinfo {author} {\bibfnamefont {L.}~\bibnamefont {Meng}}, \bibinfo
  {author} {\bibfnamefont {L.-Y.}\ \bibnamefont {Xiao}}, \bibinfo {author}
  {\bibfnamefont {M.}~\bibnamefont {Oka}}, \ and\ \bibinfo {author}
  {\bibfnamefont {S.-L.}\ \bibnamefont {Zhu}},\ }\href {\doibase
  10.1140/epjc/s10052-021-08978-0} {\bibfield  {journal} {\bibinfo  {journal}
  {Eur. Phys. J. C}\ }\textbf {\bibinfo {volume} {81}},\ \bibinfo {pages} {188}
  (\bibinfo {year} {2021})},\ \Eprint {http://arxiv.org/abs/2010.09395}
  {arXiv:2010.09395 [hep-ph]} \BibitemShut {NoStop}%
\bibitem [{\citenamefont {Xiao}\ \emph {et~al.}(2021)\citenamefont {Xiao},
  \citenamefont {Chen}, \citenamefont {Dong},\ and\ \citenamefont
  {Meng}}]{Xiao:2020ltm}%
  \BibitemOpen
  \bibfield  {author} {\bibinfo {author} {\bibfnamefont {C.-J.}\ \bibnamefont
  {Xiao}}, \bibinfo {author} {\bibfnamefont {D.-Y.}\ \bibnamefont {Chen}},
  \bibinfo {author} {\bibfnamefont {Y.-B.}\ \bibnamefont {Dong}}, \ and\
  \bibinfo {author} {\bibfnamefont {G.-W.}\ \bibnamefont {Meng}},\ }\href
  {\doibase 10.1103/PhysRevD.103.034004} {\bibfield  {journal} {\bibinfo
  {journal} {Phys. Rev. D}\ }\textbf {\bibinfo {volume} {103}},\ \bibinfo
  {pages} {034004} (\bibinfo {year} {2021})},\ \Eprint
  {http://arxiv.org/abs/2009.14538} {arXiv:2009.14538 [hep-ph]} \BibitemShut
  {NoStop}%
\bibitem [{\citenamefont {Molina}\ \emph {et~al.}(2010)\citenamefont {Molina},
  \citenamefont {Branz},\ and\ \citenamefont {Oset}}]{Molina:2010tx}%
  \BibitemOpen
  \bibfield  {author} {\bibinfo {author} {\bibfnamefont {R.}~\bibnamefont
  {Molina}}, \bibinfo {author} {\bibfnamefont {T.}~\bibnamefont {Branz}}, \
  and\ \bibinfo {author} {\bibfnamefont {E.}~\bibnamefont {Oset}},\ }\href
  {\doibase 10.1103/PhysRevD.82.014010} {\bibfield  {journal} {\bibinfo
  {journal} {Phys. Rev. D}\ }\textbf {\bibinfo {volume} {82}},\ \bibinfo
  {pages} {014010} (\bibinfo {year} {2010})},\ \Eprint
  {http://arxiv.org/abs/1005.0335} {arXiv:1005.0335 [hep-ph]} \BibitemShut
  {NoStop}%
\bibitem [{\citenamefont {Agaev}\ \emph {et~al.}(2021)\citenamefont {Agaev},
  \citenamefont {Azizi},\ and\ \citenamefont {Sundu}}]{Agaev:2021knl}%
  \BibitemOpen
  \bibfield  {author} {\bibinfo {author} {\bibfnamefont {S.~S.}\ \bibnamefont
  {Agaev}}, \bibinfo {author} {\bibfnamefont {K.}~\bibnamefont {Azizi}}, \ and\
  \bibinfo {author} {\bibfnamefont {H.}~\bibnamefont {Sundu}},\ }\href@noop {}
  {\  (\bibinfo {year} {2021})},\ \Eprint {http://arxiv.org/abs/2103.06151}
  {arXiv:2103.06151 [hep-ph]} \BibitemShut {NoStop}%
\bibitem [{\citenamefont {Tan}\ and\ \citenamefont {Ping}(2020)}]{Tan:2020cpu}%
  \BibitemOpen
  \bibfield  {author} {\bibinfo {author} {\bibfnamefont {Y.}~\bibnamefont
  {Tan}}\ and\ \bibinfo {author} {\bibfnamefont {J.}~\bibnamefont {Ping}},\
  }\href@noop {} {\  (\bibinfo {year} {2020})},\ \Eprint
  {http://arxiv.org/abs/2010.04045} {arXiv:2010.04045 [hep-ph]} \BibitemShut
  {NoStop}%
\bibitem [{\citenamefont {Qi}\ \emph {et~al.}(2021)\citenamefont {Qi},
  \citenamefont {Wang}, \citenamefont {Zhang},\ and\ \citenamefont
  {Guo}}]{Qi:2021iyv}%
  \BibitemOpen
  \bibfield  {author} {\bibinfo {author} {\bibfnamefont {J.-J.}\ \bibnamefont
  {Qi}}, \bibinfo {author} {\bibfnamefont {Z.-Y.}\ \bibnamefont {Wang}},
  \bibinfo {author} {\bibfnamefont {Z.-F.}\ \bibnamefont {Zhang}}, \ and\
  \bibinfo {author} {\bibfnamefont {X.-H.}\ \bibnamefont {Guo}},\ }\href@noop
  {} {\  (\bibinfo {year} {2021})},\ \Eprint {http://arxiv.org/abs/2101.06688}
  {arXiv:2101.06688 [hep-ph]} \BibitemShut {NoStop}%
\bibitem [{\citenamefont {Choi}\ \emph {et~al.}(2003)\citenamefont {Choi} \emph
  {et~al.}}]{Choi:2003ue}%
  \BibitemOpen
  \bibfield  {author} {\bibinfo {author} {\bibfnamefont {S.}~\bibnamefont
  {Choi}} \emph {et~al.} (\bibinfo {collaboration} {Belle}),\ }\href {\doibase
  10.1103/PhysRevLett.91.262001} {\bibfield  {journal} {\bibinfo  {journal}
  {Phys. Rev. Lett.}\ }\textbf {\bibinfo {volume} {91}},\ \bibinfo {pages}
  {262001} (\bibinfo {year} {2003})},\ \Eprint
  {http://arxiv.org/abs/hep-ex/0309032} {arXiv:hep-ex/0309032} \BibitemShut
  {NoStop}%
\bibitem [{\citenamefont {Guo}\ \emph {et~al.}(2018)\citenamefont {Guo},
  \citenamefont {Hanhart}, \citenamefont {Meißner}, \citenamefont {Wang},
  \citenamefont {Zhao},\ and\ \citenamefont {Zou}}]{Guo:2017jvc}%
  \BibitemOpen
  \bibfield  {author} {\bibinfo {author} {\bibfnamefont {F.-K.}\ \bibnamefont
  {Guo}}, \bibinfo {author} {\bibfnamefont {C.}~\bibnamefont {Hanhart}},
  \bibinfo {author} {\bibfnamefont {U.-G.}\ \bibnamefont {Meißner}}, \bibinfo
  {author} {\bibfnamefont {Q.}~\bibnamefont {Wang}}, \bibinfo {author}
  {\bibfnamefont {Q.}~\bibnamefont {Zhao}}, \ and\ \bibinfo {author}
  {\bibfnamefont {B.-S.}\ \bibnamefont {Zou}},\ }\href {\doibase
  10.1103/RevModPhys.90.015004} {\bibfield  {journal} {\bibinfo  {journal}
  {Rev. Mod. Phys.}\ }\textbf {\bibinfo {volume} {90}},\ \bibinfo {pages}
  {015004} (\bibinfo {year} {2018})},\ \Eprint
  {http://arxiv.org/abs/1705.00141} {arXiv:1705.00141 [hep-ph]} \BibitemShut
  {NoStop}%
\bibitem [{\citenamefont {Dong}\ \emph {et~al.}(2020)\citenamefont {Dong},
  \citenamefont {Lin},\ and\ \citenamefont {Zou}}]{Dong:2019ofp}%
  \BibitemOpen
  \bibfield  {author} {\bibinfo {author} {\bibfnamefont {X.-K.}\ \bibnamefont
  {Dong}}, \bibinfo {author} {\bibfnamefont {Y.-H.}\ \bibnamefont {Lin}}, \
  and\ \bibinfo {author} {\bibfnamefont {B.-S.}\ \bibnamefont {Zou}},\ }\href
  {\doibase 10.1103/PhysRevD.101.076003} {\bibfield  {journal} {\bibinfo
  {journal} {Phys. Rev. D}\ }\textbf {\bibinfo {volume} {101}},\ \bibinfo
  {pages} {076003} (\bibinfo {year} {2020})},\ \Eprint
  {http://arxiv.org/abs/1910.14455} {arXiv:1910.14455 [hep-ph]} \BibitemShut
  {NoStop}%
\bibitem [{\citenamefont {Dong}\ \emph {et~al.}(2021)\citenamefont {Dong},
  \citenamefont {Guo},\ and\ \citenamefont {Zou}}]{Dong:2021juy}%
  \BibitemOpen
  \bibfield  {author} {\bibinfo {author} {\bibfnamefont {X.-K.}\ \bibnamefont
  {Dong}}, \bibinfo {author} {\bibfnamefont {F.-K.}\ \bibnamefont {Guo}}, \
  and\ \bibinfo {author} {\bibfnamefont {B.-S.}\ \bibnamefont {Zou}},\ }\href
  {\doibase 10.13725/j.cnki.pip.2021.02.001} {\bibfield  {journal} {\bibinfo
  {journal} {Progr. Phys.}\ }\textbf {\bibinfo {volume} {41}},\ \bibinfo
  {pages} {65} (\bibinfo {year} {2021})},\ \Eprint
  {http://arxiv.org/abs/2101.01021} {arXiv:2101.01021 [hep-ph]} \BibitemShut
  {NoStop}%
\bibitem [{\citenamefont {Aubert}\ \emph {et~al.}(2003)\citenamefont {Aubert}
  \emph {et~al.}}]{Aubert:2003fg}%
  \BibitemOpen
  \bibfield  {author} {\bibinfo {author} {\bibfnamefont {B.}~\bibnamefont
  {Aubert}} \emph {et~al.} (\bibinfo {collaboration} {BaBar}),\ }\href
  {\doibase 10.1103/PhysRevLett.90.242001} {\bibfield  {journal} {\bibinfo
  {journal} {Phys. Rev. Lett.}\ }\textbf {\bibinfo {volume} {90}},\ \bibinfo
  {pages} {242001} (\bibinfo {year} {2003})},\ \Eprint
  {http://arxiv.org/abs/hep-ex/0304021} {arXiv:hep-ex/0304021} \BibitemShut
  {NoStop}%
\bibitem [{\citenamefont {Besson}\ \emph {et~al.}(2003)\citenamefont {Besson}
  \emph {et~al.}}]{Besson:2003cp}%
  \BibitemOpen
  \bibfield  {author} {\bibinfo {author} {\bibfnamefont {D.}~\bibnamefont
  {Besson}} \emph {et~al.} (\bibinfo {collaboration} {CLEO}),\ }\href {\doibase
  10.1103/PhysRevD.68.032002} {\bibfield  {journal} {\bibinfo  {journal} {Phys.
  Rev. D}\ }\textbf {\bibinfo {volume} {68}},\ \bibinfo {pages} {032002}
  (\bibinfo {year} {2003})},\ \bibinfo {note} {[Erratum: Phys.Rev.D 75, 119908
  (2007)]},\ \Eprint {http://arxiv.org/abs/hep-ex/0305100}
  {arXiv:hep-ex/0305100} \BibitemShut {NoStop}%
\bibitem [{\citenamefont {Chen}\ \emph {et~al.}(2017)\citenamefont {Chen},
  \citenamefont {Chen}, \citenamefont {Liu}, \citenamefont {Liu},\ and\
  \citenamefont {Zhu}}]{Chen:2016spr}%
  \BibitemOpen
  \bibfield  {author} {\bibinfo {author} {\bibfnamefont {H.-X.}\ \bibnamefont
  {Chen}}, \bibinfo {author} {\bibfnamefont {W.}~\bibnamefont {Chen}}, \bibinfo
  {author} {\bibfnamefont {X.}~\bibnamefont {Liu}}, \bibinfo {author}
  {\bibfnamefont {Y.-R.}\ \bibnamefont {Liu}}, \ and\ \bibinfo {author}
  {\bibfnamefont {S.-L.}\ \bibnamefont {Zhu}},\ }\href {\doibase
  10.1088/1361-6633/aa6420} {\bibfield  {journal} {\bibinfo  {journal} {Rept.
  Prog. Phys.}\ }\textbf {\bibinfo {volume} {80}},\ \bibinfo {pages} {076201}
  (\bibinfo {year} {2017})},\ \Eprint {http://arxiv.org/abs/1609.08928}
  {arXiv:1609.08928 [hep-ph]} \BibitemShut {NoStop}%
\bibitem [{\citenamefont {Liu}\ \emph {et~al.}(2013)\citenamefont {Liu},
  \citenamefont {Orginos}, \citenamefont {Guo}, \citenamefont {Hanhart},\ and\
  \citenamefont {Meissner}}]{Liu:2012zya}%
  \BibitemOpen
  \bibfield  {author} {\bibinfo {author} {\bibfnamefont {L.}~\bibnamefont
  {Liu}}, \bibinfo {author} {\bibfnamefont {K.}~\bibnamefont {Orginos}},
  \bibinfo {author} {\bibfnamefont {F.-K.}\ \bibnamefont {Guo}}, \bibinfo
  {author} {\bibfnamefont {C.}~\bibnamefont {Hanhart}}, \ and\ \bibinfo
  {author} {\bibfnamefont {U.-G.}\ \bibnamefont {Meissner}},\ }\href {\doibase
  10.1103/PhysRevD.87.014508} {\bibfield  {journal} {\bibinfo  {journal} {Phys.
  Rev. D}\ }\textbf {\bibinfo {volume} {87}},\ \bibinfo {pages} {014508}
  (\bibinfo {year} {2013})},\ \Eprint {http://arxiv.org/abs/1208.4535}
  {arXiv:1208.4535 [hep-lat]} \BibitemShut {NoStop}%
\bibitem [{\citenamefont {Martínez~Torres}\ \emph {et~al.}(2015)\citenamefont
  {Martínez~Torres}, \citenamefont {Oset}, \citenamefont {Prelovsek},\ and\
  \citenamefont {Ramos}}]{Torres:2014vna}%
  \BibitemOpen
  \bibfield  {author} {\bibinfo {author} {\bibfnamefont {A.}~\bibnamefont
  {Martínez~Torres}}, \bibinfo {author} {\bibfnamefont {E.}~\bibnamefont
  {Oset}}, \bibinfo {author} {\bibfnamefont {S.}~\bibnamefont {Prelovsek}}, \
  and\ \bibinfo {author} {\bibfnamefont {A.}~\bibnamefont {Ramos}},\ }\href
  {\doibase 10.1007/JHEP05(2015)153} {\bibfield  {journal} {\bibinfo  {journal}
  {JHEP}\ }\textbf {\bibinfo {volume} {05}},\ \bibinfo {pages} {153} (\bibinfo
  {year} {2015})},\ \Eprint {http://arxiv.org/abs/1412.1706} {arXiv:1412.1706
  [hep-lat]} \BibitemShut {NoStop}%
\bibitem [{\citenamefont {Bali}\ \emph {et~al.}(2017)\citenamefont {Bali},
  \citenamefont {Collins}, \citenamefont {Cox},\ and\ \citenamefont
  {Schäfer}}]{Bali:2017pdv}%
  \BibitemOpen
  \bibfield  {author} {\bibinfo {author} {\bibfnamefont {G.~S.}\ \bibnamefont
  {Bali}}, \bibinfo {author} {\bibfnamefont {S.}~\bibnamefont {Collins}},
  \bibinfo {author} {\bibfnamefont {A.}~\bibnamefont {Cox}}, \ and\ \bibinfo
  {author} {\bibfnamefont {A.}~\bibnamefont {Schäfer}},\ }\href {\doibase
  10.1103/PhysRevD.96.074501} {\bibfield  {journal} {\bibinfo  {journal} {Phys.
  Rev. D}\ }\textbf {\bibinfo {volume} {96}},\ \bibinfo {pages} {074501}
  (\bibinfo {year} {2017})},\ \Eprint {http://arxiv.org/abs/1706.01247}
  {arXiv:1706.01247 [hep-lat]} \BibitemShut {NoStop}%
\bibitem [{\citenamefont {Cheung}\ \emph {et~al.}(2021)\citenamefont {Cheung},
  \citenamefont {Thomas}, \citenamefont {Wilson}, \citenamefont {Moir},
  \citenamefont {Peardon},\ and\ \citenamefont {Ryan}}]{Cheung:2020mql}%
  \BibitemOpen
  \bibfield  {author} {\bibinfo {author} {\bibfnamefont {G.~K.~C.}\
  \bibnamefont {Cheung}}, \bibinfo {author} {\bibfnamefont {C.~E.}\
  \bibnamefont {Thomas}}, \bibinfo {author} {\bibfnamefont {D.~J.}\
  \bibnamefont {Wilson}}, \bibinfo {author} {\bibfnamefont {G.}~\bibnamefont
  {Moir}}, \bibinfo {author} {\bibfnamefont {M.}~\bibnamefont {Peardon}}, \
  and\ \bibinfo {author} {\bibfnamefont {S.~M.}\ \bibnamefont {Ryan}} (\bibinfo
  {collaboration} {Hadron Spectrum}),\ }\href {\doibase
  10.1007/JHEP02(2021)100} {\bibfield  {journal} {\bibinfo  {journal} {JHEP}\
  }\textbf {\bibinfo {volume} {02}},\ \bibinfo {pages} {100} (\bibinfo {year}
  {2021})},\ \Eprint {http://arxiv.org/abs/2008.06432} {arXiv:2008.06432
  [hep-lat]} \BibitemShut {NoStop}%
\bibitem [{\citenamefont {Guo}\ and\ \citenamefont
  {Meissner}(2011)}]{Guo:2011dd}%
  \BibitemOpen
  \bibfield  {author} {\bibinfo {author} {\bibfnamefont {F.-K.}\ \bibnamefont
  {Guo}}\ and\ \bibinfo {author} {\bibfnamefont {U.-G.}\ \bibnamefont
  {Meissner}},\ }\href {\doibase 10.1103/PhysRevD.84.014013} {\bibfield
  {journal} {\bibinfo  {journal} {Phys. Rev. D}\ }\textbf {\bibinfo {volume}
  {84}},\ \bibinfo {pages} {014013} (\bibinfo {year} {2011})},\ \Eprint
  {http://arxiv.org/abs/1102.3536} {arXiv:1102.3536 [hep-ph]} \BibitemShut
  {NoStop}%
\bibitem [{\citenamefont {Casalbuoni}\ \emph {et~al.}(1997)\citenamefont
  {Casalbuoni}, \citenamefont {Deandrea}, \citenamefont {Di~Bartolomeo},
  \citenamefont {Gatto}, \citenamefont {Feruglio},\ and\ \citenamefont
  {Nardulli}}]{Casalbuoni:1996pg}%
  \BibitemOpen
  \bibfield  {author} {\bibinfo {author} {\bibfnamefont {R.}~\bibnamefont
  {Casalbuoni}}, \bibinfo {author} {\bibfnamefont {A.}~\bibnamefont
  {Deandrea}}, \bibinfo {author} {\bibfnamefont {N.}~\bibnamefont
  {Di~Bartolomeo}}, \bibinfo {author} {\bibfnamefont {R.}~\bibnamefont
  {Gatto}}, \bibinfo {author} {\bibfnamefont {F.}~\bibnamefont {Feruglio}}, \
  and\ \bibinfo {author} {\bibfnamefont {G.}~\bibnamefont {Nardulli}},\ }\href
  {\doibase 10.1016/S0370-1573(96)00027-0} {\bibfield  {journal} {\bibinfo
  {journal} {Phys. Rept.}\ }\textbf {\bibinfo {volume} {281}},\ \bibinfo
  {pages} {145} (\bibinfo {year} {1997})},\ \Eprint
  {http://arxiv.org/abs/hep-ph/9605342} {arXiv:hep-ph/9605342 [hep-ph]}
  \BibitemShut {NoStop}%
\bibitem [{\citenamefont {Ding}(2009)}]{Ding:2008gr}%
  \BibitemOpen
  \bibfield  {author} {\bibinfo {author} {\bibfnamefont {G.-J.}\ \bibnamefont
  {Ding}},\ }\href {\doibase 10.1103/PhysRevD.79.014001} {\bibfield  {journal}
  {\bibinfo  {journal} {Phys. Rev. D}\ }\textbf {\bibinfo {volume} {79}},\
  \bibinfo {pages} {014001} (\bibinfo {year} {2009})},\ \Eprint
  {http://arxiv.org/abs/0809.4818} {arXiv:0809.4818 [hep-ph]} \BibitemShut
  {NoStop}%
\bibitem [{\citenamefont {Bando}\ \emph {et~al.}(1988)\citenamefont {Bando},
  \citenamefont {Kugo},\ and\ \citenamefont {Yamawaki}}]{Bando:1987br}%
  \BibitemOpen
  \bibfield  {author} {\bibinfo {author} {\bibfnamefont {M.}~\bibnamefont
  {Bando}}, \bibinfo {author} {\bibfnamefont {T.}~\bibnamefont {Kugo}}, \ and\
  \bibinfo {author} {\bibfnamefont {K.}~\bibnamefont {Yamawaki}},\ }\href
  {\doibase 10.1016/0370-1573(88)90019-1} {\bibfield  {journal} {\bibinfo
  {journal} {Phys. Rept.}\ }\textbf {\bibinfo {volume} {164}},\ \bibinfo
  {pages} {217} (\bibinfo {year} {1988})}\BibitemShut {NoStop}%
\bibitem [{\citenamefont {Isola}\ \emph {et~al.}(2003)\citenamefont {Isola},
  \citenamefont {Ladisa}, \citenamefont {Nardulli},\ and\ \citenamefont
  {Santorelli}}]{Isola:2003fh}%
  \BibitemOpen
  \bibfield  {author} {\bibinfo {author} {\bibfnamefont {C.}~\bibnamefont
  {Isola}}, \bibinfo {author} {\bibfnamefont {M.}~\bibnamefont {Ladisa}},
  \bibinfo {author} {\bibfnamefont {G.}~\bibnamefont {Nardulli}}, \ and\
  \bibinfo {author} {\bibfnamefont {P.}~\bibnamefont {Santorelli}},\ }\href
  {\doibase 10.1103/PhysRevD.68.114001} {\bibfield  {journal} {\bibinfo
  {journal} {Phys. Rev.}\ }\textbf {\bibinfo {volume} {D68}},\ \bibinfo {pages}
  {114001} (\bibinfo {year} {2003})},\ \Eprint
  {http://arxiv.org/abs/hep-ph/0307367} {arXiv:hep-ph/0307367 [hep-ph]}
  \BibitemShut {NoStop}%
\bibitem [{\citenamefont {Burakovsky}\ and\ \citenamefont
  {Goldman}(1997)}]{Burakovsky:1997dd}%
  \BibitemOpen
  \bibfield  {author} {\bibinfo {author} {\bibfnamefont {L.}~\bibnamefont
  {Burakovsky}}\ and\ \bibinfo {author} {\bibfnamefont {J.~T.}\ \bibnamefont
  {Goldman}},\ }\href {\doibase 10.1103/PhysRevD.56.R1368} {\bibfield
  {journal} {\bibinfo  {journal} {Phys. Rev.}\ }\textbf {\bibinfo {volume}
  {D56}},\ \bibinfo {pages} {R1368} (\bibinfo {year} {1997})},\ \Eprint
  {http://arxiv.org/abs/hep-ph/9703274} {arXiv:hep-ph/9703274 [hep-ph]}
  \BibitemShut {NoStop}%
\bibitem [{\citenamefont {Suzuki}(1993)}]{Suzuki:1993yc}%
  \BibitemOpen
  \bibfield  {author} {\bibinfo {author} {\bibfnamefont {M.}~\bibnamefont
  {Suzuki}},\ }\href {\doibase 10.1103/PhysRevD.47.1252} {\bibfield  {journal}
  {\bibinfo  {journal} {Phys. Rev.}\ }\textbf {\bibinfo {volume} {D47}},\
  \bibinfo {pages} {1252} (\bibinfo {year} {1993})}\BibitemShut {NoStop}%
\bibitem [{\citenamefont {Cheng}(2003)}]{Cheng:2003bn}%
  \BibitemOpen
  \bibfield  {author} {\bibinfo {author} {\bibfnamefont {H.-Y.}\ \bibnamefont
  {Cheng}},\ }\href {\doibase 10.1103/PhysRevD.67.094007} {\bibfield  {journal}
  {\bibinfo  {journal} {Phys. Rev.}\ }\textbf {\bibinfo {volume} {D67}},\
  \bibinfo {pages} {094007} (\bibinfo {year} {2003})},\ \Eprint
  {http://arxiv.org/abs/hep-ph/0301198} {arXiv:hep-ph/0301198 [hep-ph]}
  \BibitemShut {NoStop}%
\bibitem [{\citenamefont {Yang}(2011)}]{Yang:2010ah}%
  \BibitemOpen
  \bibfield  {author} {\bibinfo {author} {\bibfnamefont {K.-C.}\ \bibnamefont
  {Yang}},\ }\href {\doibase 10.1103/PhysRevD.84.034035} {\bibfield  {journal}
  {\bibinfo  {journal} {Phys. Rev.}\ }\textbf {\bibinfo {volume} {D84}},\
  \bibinfo {pages} {034035} (\bibinfo {year} {2011})},\ \Eprint
  {http://arxiv.org/abs/1011.6113} {arXiv:1011.6113 [hep-ph]} \BibitemShut
  {NoStop}%
\bibitem [{\citenamefont {Hatanaka}\ and\ \citenamefont
  {Yang}(2008)}]{Hatanaka:2008xj}%
  \BibitemOpen
  \bibfield  {author} {\bibinfo {author} {\bibfnamefont {H.}~\bibnamefont
  {Hatanaka}}\ and\ \bibinfo {author} {\bibfnamefont {K.-C.}\ \bibnamefont
  {Yang}},\ }\href {\doibase 10.1103/PhysRevD.77.094023,
  10.1103/PhysRevD.78.059902} {\bibfield  {journal} {\bibinfo  {journal} {Phys.
  Rev.}\ }\textbf {\bibinfo {volume} {D77}},\ \bibinfo {pages} {094023}
  (\bibinfo {year} {2008})},\ \bibinfo {note} {[Erratum: Phys.
  Rev.D78,059902(2008)]},\ \Eprint {http://arxiv.org/abs/0804.3198}
  {arXiv:0804.3198 [hep-ph]} \BibitemShut {NoStop}%
\bibitem [{\citenamefont {Tayduganov}\ \emph {et~al.}(2012)\citenamefont
  {Tayduganov}, \citenamefont {Kou},\ and\ \citenamefont
  {Le~Yaouanc}}]{Tayduganov:2011ui}%
  \BibitemOpen
  \bibfield  {author} {\bibinfo {author} {\bibfnamefont {A.}~\bibnamefont
  {Tayduganov}}, \bibinfo {author} {\bibfnamefont {E.}~\bibnamefont {Kou}}, \
  and\ \bibinfo {author} {\bibfnamefont {A.}~\bibnamefont {Le~Yaouanc}},\
  }\href {\doibase 10.1103/PhysRevD.85.074011} {\bibfield  {journal} {\bibinfo
  {journal} {Phys. Rev.}\ }\textbf {\bibinfo {volume} {D85}},\ \bibinfo {pages}
  {074011} (\bibinfo {year} {2012})},\ \Eprint {http://arxiv.org/abs/1111.6307}
  {arXiv:1111.6307 [hep-ph]} \BibitemShut {NoStop}%
\bibitem [{\citenamefont {Divotgey}\ \emph {et~al.}(2013)\citenamefont
  {Divotgey}, \citenamefont {Olbrich},\ and\ \citenamefont
  {Giacosa}}]{Divotgey:2013jba}%
  \BibitemOpen
  \bibfield  {author} {\bibinfo {author} {\bibfnamefont {F.}~\bibnamefont
  {Divotgey}}, \bibinfo {author} {\bibfnamefont {L.}~\bibnamefont {Olbrich}}, \
  and\ \bibinfo {author} {\bibfnamefont {F.}~\bibnamefont {Giacosa}},\ }\href
  {\doibase 10.1140/epja/i2013-13135-3} {\bibfield  {journal} {\bibinfo
  {journal} {Eur. Phys. J.}\ }\textbf {\bibinfo {volume} {A49}},\ \bibinfo
  {pages} {135} (\bibinfo {year} {2013})},\ \Eprint
  {http://arxiv.org/abs/1306.1193} {arXiv:1306.1193 [hep-ph]} \BibitemShut
  {NoStop}%
\bibitem [{\citenamefont {Zhang}\ \emph {et~al.}(2018)\citenamefont {Zhang},
  \citenamefont {Guo},\ and\ \citenamefont {Wang}}]{Zhang:2017cbi}%
  \BibitemOpen
  \bibfield  {author} {\bibinfo {author} {\bibfnamefont {Z.-Q.}\ \bibnamefont
  {Zhang}}, \bibinfo {author} {\bibfnamefont {H.}~\bibnamefont {Guo}}, \ and\
  \bibinfo {author} {\bibfnamefont {S.-Y.}\ \bibnamefont {Wang}},\ }\href
  {\doibase 10.1140/epjc/s10052-018-5674-7} {\bibfield  {journal} {\bibinfo
  {journal} {Eur. Phys. J.}\ }\textbf {\bibinfo {volume} {C78}},\ \bibinfo
  {pages} {219} (\bibinfo {year} {2018})},\ \Eprint
  {http://arxiv.org/abs/1705.00524} {arXiv:1705.00524 [hep-ph]} \BibitemShut
  {NoStop}%
\bibitem [{\citenamefont {Roca}\ \emph {et~al.}(2005)\citenamefont {Roca},
  \citenamefont {Oset},\ and\ \citenamefont {Singh}}]{Roca:2005nm}%
  \BibitemOpen
  \bibfield  {author} {\bibinfo {author} {\bibfnamefont {L.}~\bibnamefont
  {Roca}}, \bibinfo {author} {\bibfnamefont {E.}~\bibnamefont {Oset}}, \ and\
  \bibinfo {author} {\bibfnamefont {J.}~\bibnamefont {Singh}},\ }\href
  {\doibase 10.1103/PhysRevD.72.014002} {\bibfield  {journal} {\bibinfo
  {journal} {Phys. Rev.}\ }\textbf {\bibinfo {volume} {D72}},\ \bibinfo {pages}
  {014002} (\bibinfo {year} {2005})},\ \Eprint
  {http://arxiv.org/abs/hep-ph/0503273} {arXiv:hep-ph/0503273 [hep-ph]}
  \BibitemShut {NoStop}%
\bibitem [{\citenamefont {Geng}\ \emph {et~al.}(2007)\citenamefont {Geng},
  \citenamefont {Oset}, \citenamefont {Roca},\ and\ \citenamefont
  {Oller}}]{Geng:2006yb}%
  \BibitemOpen
  \bibfield  {author} {\bibinfo {author} {\bibfnamefont {L.~S.}\ \bibnamefont
  {Geng}}, \bibinfo {author} {\bibfnamefont {E.}~\bibnamefont {Oset}}, \bibinfo
  {author} {\bibfnamefont {L.}~\bibnamefont {Roca}}, \ and\ \bibinfo {author}
  {\bibfnamefont {J.~A.}\ \bibnamefont {Oller}},\ }\href {\doibase
  10.1103/PhysRevD.75.014017} {\bibfield  {journal} {\bibinfo  {journal} {Phys.
  Rev.}\ }\textbf {\bibinfo {volume} {D75}},\ \bibinfo {pages} {014017}
  (\bibinfo {year} {2007})},\ \Eprint {http://arxiv.org/abs/hep-ph/0610217}
  {arXiv:hep-ph/0610217 [hep-ph]} \BibitemShut {NoStop}%
\bibitem [{\citenamefont {Wang}\ \emph {et~al.}(2019)\citenamefont {Wang},
  \citenamefont {Roca},\ and\ \citenamefont {Oset}}]{Wang:2019mph}%
  \BibitemOpen
  \bibfield  {author} {\bibinfo {author} {\bibfnamefont {G.~Y.}\ \bibnamefont
  {Wang}}, \bibinfo {author} {\bibfnamefont {L.}~\bibnamefont {Roca}}, \ and\
  \bibinfo {author} {\bibfnamefont {E.}~\bibnamefont {Oset}},\ }\href {\doibase
  10.1103/PhysRevD.100.074018} {\bibfield  {journal} {\bibinfo  {journal}
  {Phys. Rev.}\ }\textbf {\bibinfo {volume} {D100}},\ \bibinfo {pages} {074018}
  (\bibinfo {year} {2019})},\ \Eprint {http://arxiv.org/abs/1907.09188}
  {arXiv:1907.09188 [hep-ph]} \BibitemShut {NoStop}%
\bibitem [{\citenamefont {Zhang}\ \emph {et~al.}(2006)\citenamefont {Zhang},
  \citenamefont {Chiang}, \citenamefont {Shen},\ and\ \citenamefont
  {Zou}}]{Zhang:2006ix}%
  \BibitemOpen
  \bibfield  {author} {\bibinfo {author} {\bibfnamefont {Y.-J.}\ \bibnamefont
  {Zhang}}, \bibinfo {author} {\bibfnamefont {H.-C.}\ \bibnamefont {Chiang}},
  \bibinfo {author} {\bibfnamefont {P.-N.}\ \bibnamefont {Shen}}, \ and\
  \bibinfo {author} {\bibfnamefont {B.-S.}\ \bibnamefont {Zou}},\ }\href
  {\doibase 10.1103/PhysRevD.74.014013} {\bibfield  {journal} {\bibinfo
  {journal} {Phys. Rev.}\ }\textbf {\bibinfo {volume} {D74}},\ \bibinfo {pages}
  {014013} (\bibinfo {year} {2006})},\ \Eprint
  {http://arxiv.org/abs/hep-ph/0604271} {arXiv:hep-ph/0604271 [hep-ph]}
  \BibitemShut {NoStop}%
\bibitem [{\citenamefont {Weinberg}(1966)}]{Weinberg:1966kf}%
  \BibitemOpen
  \bibfield  {author} {\bibinfo {author} {\bibfnamefont {S.}~\bibnamefont
  {Weinberg}},\ }\href {\doibase 10.1103/PhysRevLett.17.616} {\bibfield
  {journal} {\bibinfo  {journal} {Phys. Rev. Lett.}\ }\textbf {\bibinfo
  {volume} {17}},\ \bibinfo {pages} {616} (\bibinfo {year} {1966})}\BibitemShut
  {NoStop}%
\bibitem [{\citenamefont {Tomozawa}(1966)}]{Tomozawa:1966jm}%
  \BibitemOpen
  \bibfield  {author} {\bibinfo {author} {\bibfnamefont {Y.}~\bibnamefont
  {Tomozawa}},\ }\href {\doibase 10.1007/BF02857517} {\bibfield  {journal}
  {\bibinfo  {journal} {Nuovo Cim. A}\ }\textbf {\bibinfo {volume} {46}},\
  \bibinfo {pages} {707} (\bibinfo {year} {1966})}\BibitemShut {NoStop}%
\bibitem [{\citenamefont {Wang}\ \emph {et~al.}(2013)\citenamefont {Wang},
  \citenamefont {Hanhart},\ and\ \citenamefont {Zhao}}]{Wang:2013cya}%
  \BibitemOpen
  \bibfield  {author} {\bibinfo {author} {\bibfnamefont {Q.}~\bibnamefont
  {Wang}}, \bibinfo {author} {\bibfnamefont {C.}~\bibnamefont {Hanhart}}, \
  and\ \bibinfo {author} {\bibfnamefont {Q.}~\bibnamefont {Zhao}},\ }\href
  {\doibase 10.1103/PhysRevLett.111.132003} {\bibfield  {journal} {\bibinfo
  {journal} {Phys. Rev. Lett.}\ }\textbf {\bibinfo {volume} {111}},\ \bibinfo
  {pages} {132003} (\bibinfo {year} {2013})},\ \Eprint
  {http://arxiv.org/abs/1303.6355} {arXiv:1303.6355 [hep-ph]} \BibitemShut
  {NoStop}%
\bibitem [{\citenamefont {Chen}\ \emph {et~al.}(2019)\citenamefont {Chen},
  \citenamefont {Dai}, \citenamefont {Guo},\ and\ \citenamefont
  {Kubis}}]{Chen:2019mgp}%
  \BibitemOpen
  \bibfield  {author} {\bibinfo {author} {\bibfnamefont {Y.-H.}\ \bibnamefont
  {Chen}}, \bibinfo {author} {\bibfnamefont {L.-Y.}\ \bibnamefont {Dai}},
  \bibinfo {author} {\bibfnamefont {F.-K.}\ \bibnamefont {Guo}}, \ and\
  \bibinfo {author} {\bibfnamefont {B.}~\bibnamefont {Kubis}},\ }\href
  {\doibase 10.1103/PhysRevD.99.074016} {\bibfield  {journal} {\bibinfo
  {journal} {Phys. Rev. D}\ }\textbf {\bibinfo {volume} {99}},\ \bibinfo
  {pages} {074016} (\bibinfo {year} {2019})},\ \Eprint
  {http://arxiv.org/abs/1902.10957} {arXiv:1902.10957 [hep-ph]} \BibitemShut
  {NoStop}%
\bibitem [{\citenamefont {Tornqvist}(1994)}]{Tornqvist:1993ng}%
  \BibitemOpen
  \bibfield  {author} {\bibinfo {author} {\bibfnamefont {N.~A.}\ \bibnamefont
  {Tornqvist}},\ }\href {\doibase 10.1007/BF01413192} {\bibfield  {journal}
  {\bibinfo  {journal} {Z. Phys.}\ }\textbf {\bibinfo {volume} {C61}},\
  \bibinfo {pages} {525} (\bibinfo {year} {1994})},\ \Eprint
  {http://arxiv.org/abs/hep-ph/9310247} {arXiv:hep-ph/9310247 [hep-ph]}
  \BibitemShut {NoStop}%
\bibitem [{\citenamefont {Weinberg}(1965)}]{Weinberg:1965zz}%
  \BibitemOpen
  \bibfield  {author} {\bibinfo {author} {\bibfnamefont {S.}~\bibnamefont
  {Weinberg}},\ }\href {\doibase 10.1103/PhysRev.137.B672} {\bibfield
  {journal} {\bibinfo  {journal} {Phys. Rev.}\ }\textbf {\bibinfo {volume}
  {137}},\ \bibinfo {pages} {B672} (\bibinfo {year} {1965})}\BibitemShut
  {NoStop}%
\bibitem [{\citenamefont {Baru}\ \emph {et~al.}(2004)\citenamefont {Baru},
  \citenamefont {Haidenbauer}, \citenamefont {Hanhart}, \citenamefont
  {Kalashnikova},\ and\ \citenamefont {Kudryavtsev}}]{Baru:2003qq}%
  \BibitemOpen
  \bibfield  {author} {\bibinfo {author} {\bibfnamefont {V.}~\bibnamefont
  {Baru}}, \bibinfo {author} {\bibfnamefont {J.}~\bibnamefont {Haidenbauer}},
  \bibinfo {author} {\bibfnamefont {C.}~\bibnamefont {Hanhart}}, \bibinfo
  {author} {\bibfnamefont {Y.}~\bibnamefont {Kalashnikova}}, \ and\ \bibinfo
  {author} {\bibfnamefont {A.~E.}\ \bibnamefont {Kudryavtsev}},\ }\href
  {\doibase 10.1016/j.physletb.2004.01.088} {\bibfield  {journal} {\bibinfo
  {journal} {Phys. Lett. B}\ }\textbf {\bibinfo {volume} {586}},\ \bibinfo
  {pages} {53} (\bibinfo {year} {2004})},\ \Eprint
  {http://arxiv.org/abs/hep-ph/0308129} {arXiv:hep-ph/0308129} \BibitemShut
  {NoStop}%
\bibitem [{\citenamefont {Casalbuoni}\ \emph {et~al.}(1993)\citenamefont
  {Casalbuoni}, \citenamefont {Deandrea}, \citenamefont {Di~Bartolomeo},
  \citenamefont {Gatto}, \citenamefont {Feruglio},\ and\ \citenamefont
  {Nardulli}}]{Casalbuoni:1992dx}%
  \BibitemOpen
  \bibfield  {author} {\bibinfo {author} {\bibfnamefont {R.}~\bibnamefont
  {Casalbuoni}}, \bibinfo {author} {\bibfnamefont {A.}~\bibnamefont
  {Deandrea}}, \bibinfo {author} {\bibfnamefont {N.}~\bibnamefont
  {Di~Bartolomeo}}, \bibinfo {author} {\bibfnamefont {R.}~\bibnamefont
  {Gatto}}, \bibinfo {author} {\bibfnamefont {F.}~\bibnamefont {Feruglio}}, \
  and\ \bibinfo {author} {\bibfnamefont {G.}~\bibnamefont {Nardulli}},\ }\href
  {\doibase 10.1016/0370-2693(93)90895-O} {\bibfield  {journal} {\bibinfo
  {journal} {Phys. Lett. B}\ }\textbf {\bibinfo {volume} {299}},\ \bibinfo
  {pages} {139} (\bibinfo {year} {1993})},\ \Eprint
  {http://arxiv.org/abs/hep-ph/9211248} {arXiv:hep-ph/9211248} \BibitemShut
  {NoStop}%
\bibitem [{\citenamefont {Jing}\ \emph {et~al.}(2020)\citenamefont {Jing},
  \citenamefont {Shen},\ and\ \citenamefont {Guo}}]{Jing:2020tth}%
  \BibitemOpen
  \bibfield  {author} {\bibinfo {author} {\bibfnamefont {H.-J.}\ \bibnamefont
  {Jing}}, \bibinfo {author} {\bibfnamefont {C.-W.}\ \bibnamefont {Shen}}, \
  and\ \bibinfo {author} {\bibfnamefont {F.-K.}\ \bibnamefont {Guo}},\
  }\href@noop {} {\  (\bibinfo {year} {2020})},\ \Eprint
  {http://arxiv.org/abs/2005.01942} {arXiv:2005.01942 [hep-ph]} \BibitemShut
  {NoStop}%
\bibitem [{\citenamefont {Close}(1979)}]{Close:1979}%
  \BibitemOpen
  \bibfield  {author} {\bibinfo {author} {\bibfnamefont {F.~E.}\ \bibnamefont
  {Close}},\ }\href
  {http://gen.lib.rus.ec/book/index.php?md5=e509303f62bb87d49256a42f36b94930}
  {\emph {\bibinfo {title} {An introduction to quarks and partons}}}\ (\bibinfo
   {publisher} {Academic Press},\ \bibinfo {year} {1979})\BibitemShut {NoStop}%
\end{thebibliography}%

\end{document}